\documentclass[12pt,preprint]{aastex}

\newcommand{\xmmnewton}{{\it XMM-Newton}}
\newcommand{\rosat}{{\it ROSAT}}

\newcommand{\einstein}{{\it Einstein}}
\newcommand{\chandra}{{\it Chandra}}

\newcommand{\swift}{{\it Swift}}
\newcommand{\galex}{{\it Galex}}
\newcommand{\spitzer}{{\it Spitzer}}

\newcommand{\NH}{\mbox {$N_{\rm H}$}}
\newcommand{\nh}{\mbox {$N_{\rm H}$}}

\newcommand{\hi}{H\,{\sc i}}
\newcommand{\hii}{H\,{\sc ii}}
\newcommand{\sii}{S\,{\sc ii}}

\newcommand{\oiii}{O\,{\sc iii}}

\newcommand{\about}{$\sim$\kern.03em}
\newcommand{\erg}{\,erg s$^{-1}$}
\newcommand{\ergs}[1]{$\times 10^{#1}$ erg s$^{-1}$}
\newcommand{\ergscms}[1]{$\times 10^{#1}$ erg cm$^{-2}$ s$^{-1}$}
\newcommand{\chase}{ChASeM33}
\newcommand{\m}{M\,33}
\newcommand{\wa}{{\tt wavdetect}}
\newcommand{\ace}{{\tt AE}}
\newcommand{\halpha}{${\rm H}\alpha$}

\slugcomment{Draft \today}

\shorttitle{The \chase\ source catalog}
\shortauthors{T\"ullmann et al.}

\begin{document}


\title{The \chandra\/ ACIS Survey of \m\ (\chase):\\ The final source catalog}


\author{R. T\"ullmann\altaffilmark{1},
T. J. Gaetz\altaffilmark{1},
P. P. Plucinsky\altaffilmark{1},
K. D. Kuntz\altaffilmark{2,3},
B. F. Williams\altaffilmark{4},
W. Pietsch\altaffilmark{5},
F. Haberl\altaffilmark{5},
K. S. Long\altaffilmark{6},
W. P. Blair\altaffilmark{2},
M. Sasaki\altaffilmark{7},
P. F. Winkler\altaffilmark{8},
P. Challis\altaffilmark{1}, 
T. G. Pannuti\altaffilmark{9},
R. J. Edgar\altaffilmark{1}, 
D. J. Helfand\altaffilmark{10},
J. P. Hughes\altaffilmark{11}, 
R. P. Kirshner\altaffilmark{1}, 
T. Mazeh\altaffilmark{12}, and
A. Shporer\altaffilmark{13,14}
}

\altaffiltext{1}{Harvard-Smithsonian Center for Astrophysics, 60 Garden Street, Cambridge, MA 02138; rtuellmann@cfa.harvard.edu}
\altaffiltext{2}{Department of Physics and Astronomy, Johns Hopkins University, 3400 North Charles Street, Baltimore, MD 21218}
\altaffiltext{3}{NASA Goddard Space Flight Center, Code 662, Greenbelt, MD 20771} 
\altaffiltext{4}{Astronomy Department, University of Washington, Seattle, WA 98195}
\altaffiltext{5}{Max-Planck-Institut f\"ur Extraterrestrische Physik, Giessenbachstra\ss e, 85741 Garching, Germany}
\altaffiltext{6}{Space Telescope Science Institute, 3700 San Martin Drive, Baltimore, MD 21218}
\altaffiltext{7}{Institut f\"ur Astronomie und Astrophysik, Eberhard Karls Universit\"at, T\"ubingen, Germany}
\altaffiltext{8}{Department of Physics, Middlebury College,  Middlebury, VT 05753}
\altaffiltext{9}{Space Science Center, 235 Martindale Drive, Morehead State University, Morehead, KY 40351}
\altaffiltext{10}{Columbia Astrophysics Laboratory, 550 W. $120^{\mathrm{th}}$ St., New York, NY 10027}
\altaffiltext{11}{Department of Physics and Astronomy, Rutgers University, 136 Frelinghuysen Road, Piscataway, NJ 08854-8019}
\altaffiltext{12}{School of Physics and Astronomy, Raymond and Beverly Sackler Faculty of Exact Sciences, Tel Aviv University, Tel Aviv, Israel 69978}
\altaffiltext{13}{Las Cumbres Observatory Global Telescope Network, 6740 Cortona Drive, Santa Barbara, CA 93117, USA}
\altaffiltext{14}{Department of Physics, Broida Hall, University of California, Santa Barbara, CA 93106, USA}

\begin{abstract}
This study presents the final source catalog of the \chandra\/ ACIS Survey of \m\ (\chase).  With a total exposure time of 1.4\,Ms, \chase\ covers $\sim$70\% of the $D_{25}$ isophote ($R \approx 4.0$\,kpc) of \m\ and provides the deepest, most complete, and detailed look at a spiral galaxy in X-rays. The source catalog includes 662 sources, reaches a limiting unabsorbed luminosity of $\sim$\,2.4\ergs{34} in the 0.35\,--\,8.0\,keV energy band, and contains source positions, source net counts, fluxes and significances in several energy bands, and information on source variability. The analysis challenges posed by \chase\ and the techniques adopted to address these challenges are discussed. 
To constrain the nature of the detected X-ray source, hardness ratios were constructed and spectra were fit for 254 sources, followup MMT spectra of 116 sources were acquired, and cross-correlations with previous X-ray catalogs and other multi-wavelength data were generated. Based on this effort, 183 of the 662 \chase\ sources could be identified. Finally, the luminosity function for the detected point sources as well as the one for the X-ray binaries in \m\ is presented. mk78The luminosity functions in the soft band (0.5-2.0 keV) and the hard band (2.0-8.0 keV) have a limiting luminosity at the 90\% completeness limit of $4.0\times10^{34}$\erg\ and $1.6\times10^{35}$\erg (for $D$\,=\,817\,kpc), respectively, which is significantly lower than what was reported by previous X-ray binary population studies in galaxies more distant than M33. The resulting distribution is consistent with a dominant population of high mass X-ray binaries as would be expected for M33.
\end{abstract}


\section{Introduction}
The \chandra\ ACIS survey of M33 (ChASeM33) is a very deep X-ray survey of \m, the nearest late-type face-on spiral galaxy, comprising a total of 1.4\,Ms of observing time. \m\ is located at a distance of about 817\,kpc \citep{freed01} and is viewed at an intermediate inclination angle of $i=56^{\circ}\pm1^{\circ}$ \citep{zar89}. The corresponding Galactic foreground column density is $\nh=6.0\times10^{20}$\,cm$^{-2}$ \citep{dickey90} which is relatively low. These properties make \m\ a prime target to exploit \chandra's high spatial resolution to gain new insights on the X-ray properties and the evolution of spiral galaxies. A detailed discussion of \chase, its scientific goals, the survey layout as well as the observing strategy was provided by \citet{ppp08}. They also provided initial science results based on the data collected during the first half of the observing cycle (from September 2005 through July 2006), such as the first version of the X-ray source catalog, hardness ratios of the X-ray sources, and an evaluation of the diffuse X-ray emission in the giant \hii\ region NGC\,604 and of the X-ray emitting SNRs in the southern spiral arm.

\begin{figure}[t]
\centering
\includegraphics[width=16.5cm,height=8.25cm,clip]{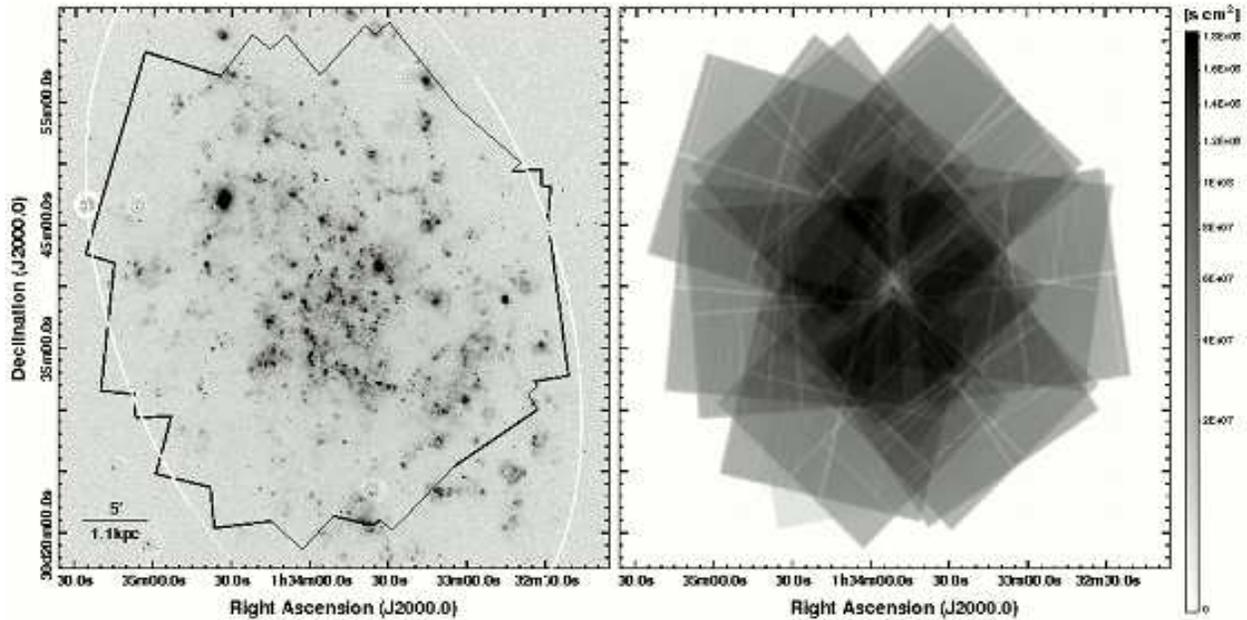}
\caption{\label{f1}
Left panel: Total coverage (black solid line) of the \chase\ and archival pointings together with the $D_{25}$ ellipse are shown on an continuum-subtracted \halpha\/ image \citep{mcneil06}. Right panel: Merged exposure map shown on a square root scale in the 0.35\,--\,8.0\,keV energy band. Only ACIS-I CCDs 0-3 have been used for the analysis.}
\end{figure}

In the present study, we make use of all \chase\ observations and two archival observations to generate the final \chase\ point source catalog of M33. The full survey covers about 70\% of the area enclosed by the $D_{25}$ B-band isophote (Fig.~\ref{f1}) and reaches a radial angular extent of $\sim$18$\arcmin$ which corresponds to $\sim$4\,kpc for the assumed distance to \m. Our survey attains an unabsorbed limiting luminosity of 2.4\ergs{34} in the 0.35\,--\,8.0\,keV energy band which makes \chase\ the deepest and spatially best resolved X-ray survey of any galaxy so far.
\begin{deluxetable}{lccc}
\tabletypesize{\scriptsize}
\tablecaption{\label{surveys}Summary of Prior X-ray Surveys of \m}
\tablewidth{0pt}
\tablehead{
\colhead{Observatory} &
\colhead{Number of Detected Sources} &
\colhead{Limiting Unabsorbed Luminosity} &
\colhead{References} \\
&
&
\colhead{(10$^{35}$ ergs/sec)\tablenotemark{a}} 
&
}
\startdata
{\it Einstein}   & 17$^b$  & --        & (1), (2), (3) \\
{\it ROSAT}      & 184$^c$ & $\sim$2.0 & (4), (5), (6) \\
{\it XMM-Newton} & 447$^d$ & $\sim$1.0 & (7), (8)      \\
{\it Chandra}    & 261$^e$ & $\sim$0.2 & (9)           \\
\enddata
\vspace{-0.75cm}
\tablecomments{$^a$For the {\it ROSAT}, {\it XMM-Newton}, and {\it Chandra} observations, the corresponding energy ranges are 0.12-2.48 keV, 0.2-4.5 keV, and 0.35-8.0 keV, respectively. $^b$Significant component of diffuse emission detected. $^c$Total number of sources located within $\sim$50' of nucleus. $^d$Total number of sources within $\sim$32$\arcmin$ of nucleus. $^e$Total number of sources in a 0.16 square degree field including the nucleus and a field northwest of the center that included NGC\,604. Summary of References: (1) -- \citet{long81}, (2) -- \citet{mara83}, (3) -- \citet{trine88}, (4) -- \citet{schube95},(5) -- \citet{long96}, (6) -- \citet{hapi01}, (7) -- \citet{pietsch04}, (8) -- \citet{misa06}, (9) -- \citet{grimm05}.} 
\end{deluxetable}

The X-ray source population of \m\ was first surveyed by the \einstein\ X-ray observatory \citep{long81,mara83,trine88}, revealing 17 sources and a significant diffuse component to the emission. Subsequent observations carried out with \rosat\ increased the number of known X-ray sources significantly \citep{schube95,long96} culminating in the catalog provided by \citet[][HP01 hereafter]{hapi01} of 184 discrete X-ray sources located within 50\arcmin\ of the nucleus to a limiting unabsorbed luminosity $L_X(0.12-2.48\,{\rm keV})\approx2\times10^{35}$\erg. The more recent \xmmnewton\ surveys from \citet[][PMH04 hereafter]{pietsch04} and \citet[][MPH06 hereafter]{misa06} cover sources located within a radius of $\sim$32$\arcmin$ from the center of the galaxy, raised the total number of sources detected within the \m\ field of view (FOV) to 447 and reached a limiting unabsorbed luminosity as low as $L_X(0.2-4.5 {\rm keV})\approx1\times10^{35}$erg\,s$^{-1}$. 
Lastly, an initial study of 261 discrete X-ray sources detected by \chandra\ towards \m\ -- for a field that covered a 0.16 square degree region including the center of the galaxy (ObsIDs 786, and 1730) and a field northwest of the center which covers the giant \hii\ region NGC604 (ObsID 2023) -- was provided by \citet[][G05 hereafter]{grimm05}. The limiting unabsorbed luminosity claimed by G05 is $L_X(0.35-8.0 keV)\approx2\times10^{34}$ erg\,s$^{-1}$ which is comparable to our survey's limit, because G05 used a lower significance threshold for their source detection. A brief overview of the previous X-ray surveys of \m\ is presented in Table~\ref{surveys}.

The limiting luminosity attained by \chase\ is significantly lower than that for comparable surveys of other galaxies, such as the \rosat\ surveys of the Local Group galaxy M31 \citep{cappa89,supper01}, the Small Magellanic Cloud \citep[SMC,][]{kaha99} and the \chandra\ surveys of spiral galaxies located outside of the Local Group, such as M101 \citep{pence01,kuntz10}.

For all of these surveys, a significant number of background AGN and foreground stars were detected which contaminated the sources that are in fact native to \m, such as X-ray binaries (XRBs), supersoft sources (SSSs) or SNRs. 
Perhaps with the exception of \chandra, the relatively poor spatial resolution of the aforementioned X-ray satellites has certainly introduced a severe source confusion bias into the existing data which also hampered the cross-correlation of the X-ray sources with those observed at different wavelengths. In \m\ this is particularly true for regions where source crowding is an issue, such as the nucleus and NGC604. Therefore, for a reliable source identification which might ultimately affect the shape of the luminosity function (LF) of the discrete sources, high spatial resolution is essential.  

Before and after the publication of the \chase\ firstlook paper by \citet{ppp08} our collaboration exploited this complex data set and published results related to the primary objectives of \chase. These are: (1) an investigation of the SNR population in \m\ \citep{gaetz07,long10}, (2) the establishment of the morphology and physical parameters of extended sources such as the giant \hii-regions NGC\,604, and IC\,131 as well as the large-scale hot gas in the spiral arms \citep{tull08,tull09, kuntzea11}, (3) the detection and analysis of X-ray binaries \citep{pietsch06,pietsch09} and transient sources \citep{ben08}, and (4) the detection of all X-ray sources down to the sensitivity limit and the location of possible counterparts including stars and background AGN \citep{ppp08}. The searches for such counterparts help us to identify and properly classify X-ray sources unassociated with \m, which in turn allows us to construct the intrinsic X-ray LF of \m. In the present paper, we concentrate on science objective (4). 

The organization of this paper is as follows. In Section \ref{sec-obs_dared} we provide detailed information on data reduction and on the generation of the final source catalog. Although this paper concentrates on the discrete X-ray sources in \m, we also discuss the criteria which need to be met to classify a source as ``extended'' (Sect.~\ref{subsec-es}).
In Section \ref{sec-resdis} the main results of our survey are presented and discussed, covering topics ranging from basic source properties (\ref{sec-srccat}), a source cross-correlation with other catalogs (\ref{sec-comp}), hardness ratios (\ref{sec-hardness}), an analysis of time variable (\ref{subsec-variable}) and supersoft sources (\ref{sec-sss}), a spectral X-ray analysis of the 15 brightest sources (\ref{sec-specana}), to the X-ray LF and radial source distribution of \m\ (\ref{sec-lumfun}). The paper concludes with a summary and the main conclusions in Section \ref{sec-summary}.

\section{Observations and Data Reduction}\label{sec-obs_dared}

\subsection{Observations}
\chase\ was designed to provide two uninterrupted observations of the seven survey fields, each with an integration time of 100\,ks \citep[see][for details]{ppp08}. The individual observations per field were separated by 5\,--\,13 months depending on field to allow for an investigation of the time variability of the X-ray sources. In some cases, as for example for Fields 5 and 6, scheduling constraints required a larger number of shorter observations.
In addition to the \chase\ observations, we also utilized two archival ACIS-I observations of \m\ (ObsIDs 1730 and 2023), one centered on the nucleus and one centered on the giant \hii\/ region NGC\,604. The results presented here use the full survey data obtained during September 2005 to November 2006. All observations were performed with ACIS-I as the primary instrument in {\tt VFAINT} mode, except ObsID 2023 which was observed in {\tt FAINT} mode. The coverage of the survey is shown in Fig.~\ref{f1} together with the merged exposure map of the \chase\ and the archival ACIS-I observations 1730 and 2023. The exposure map was binned by 1 pixel (or 0\farcs492). A journal of observations is given in Table~\ref{obs}. 
\begin{deluxetable}{cccccccc}
\tabletypesize{\footnotesize}
\tablecaption{\label{obs} List of all \chase\ and archival observations.}
\tablehead{
\colhead{ObsID} & 
\colhead{Field} & 
\colhead{Epoch} & 
\multicolumn{2}{c}{Pointing direction} & 
\colhead{Obs. start} & 
\colhead{Roll angle} & 
\colhead{Exposure} \\
\colhead{} & 
\colhead{No.} & 
\colhead{} & 
\colhead{RA (J2000.0)} & 
\colhead{Dec (J2000.0)} & 
\colhead{date} & 
\colhead{[\degr]} & 
\colhead{(cleaned) [ks]}
} 
\startdata
6376 & 1 & e1 & 01:33:51.143 & +30:39:20.54 & 2006/03/03 & 308.48 & 94.3 \\
6377 & 1 & e2 & 01:33:50.182 & +30:39:51.29 & 2006/09/25 & 142.01 & 93.2 \\
6378 & 2 & e1 & 01:34:13.208 & +30:48:02.91 & 2005/09/21 & 140.21 & 95.5 \\
6379 & 2 & e2 & 01:34:13.468 & +30:48:04.21 & 2006/09/04 & 127.88 & 54.3 \\
7402 & 2 & e2 & 01:34:13.470 & +30:48:04.17 & 2006/09/07 & 127.88 & 45.2 \\
6380 & 3 & e1 & 01:33:33.310 & +30:48:55.50 & 2005/09/23 & 140.21 & 90.5 \\
6381 & 3 & e2 & 01:33:33.475 & +30:48:56.44 & 2006/09/12 & 132.17 & 99.6 \\
6382 & 4 & e1 & 01:33:08.204 & +30:40:10.59 & 2005/11/23 & 262.21 & 72.7 \\
7226 & 4 & e1 & 01:33:08.211 & +30:40:10.55 & 2005/11/26 & 262.21 & 25.2 \\
6383 & 4 & e2 & 01:33:09.206 & +30:40:40.97 & 2006/06/15 & \ \,97.80  & 91.9 \\
7170 & 5 & e1 & 01:33:27.204 & +30:31:39.31 & 2005/09/26 & 145.71 & 41.5 \\
7171 & 5 & e1 & 01:33:27.205 & +30:31:39.29 & 2005/09/29 & 145.71 & 38.2 \\
6384 & 5 & e1 & 01:33:27.210 & +30:31:39.25 & 2005/10/01 & 145.71 & 22.2 \\
6385 & 5 & e2 & 01:33:27.402 & +30:31:40.64 & 2006/09/18 & 135.75 & 90.4 \\ 
6386 & 6 & e1 & 01:34:06.489 & +30:30:26.69 & 2005/10/31 & 224.21 & 14.8 \\
7196 & 6 & e1 & 01:34:06.500 & +30:30:26.73 & 2005/11/02 & 224.21 & 23.0 \\
7197 & 6 & e1 & 01:34:06.501 & +30:30:26.62 & 2005/11/03 & 224.21 & 12.8 \\
7198 & 6 & e1 & 01:34:06.505 & +30:30:26.82 & 2005/11/05 & 224.21 & 22.0 \\
7199 & 6 & e1 & 01:34:06.489 & +30:30:26.78 & 2005/11/06 & 224.21 & 14.8 \\
7208 & 6 & e1 & 01:34:06.992 & +30:30:18.96 & 2005/11/21 & 259.45 & 11.7 \\
6387 & 6 & e2 & 01:34:07.920 & +30:30:49.21 & 2006/06/26 & 103.21 & 78.3 \\
7344 & 6 & e2 & 01:34:07.921 & +30:30:49.31 & 2006/07/01 & 103.21 & 21.7 \\
6388 & 7 & e1 & 01:34:33.542 & +30:39:00.27 & 2006/06/09 & \ \,94.87  & 89.7 \\
6389 & 7 & e2 & 01:34:32.547 & +30:38:29.65 & 2006/11/28 & 265.72 & 96.8 \\
\hline
1730 & \multicolumn{2}{c}{archival ObsID} & 01:33:50.953 & +30:39:57.27 & 2000/07/12 & 108.58 & 47.3 \\
2023 & \multicolumn{2}{c}{archival ObsID} & 01:34:34.630 & +30:47:58.17 & 2001/07/06 & 106.55 & 89.4 \\
\enddata
\end{deluxetable}

As can be seen from Fig.~\ref{f1}, the merged exposure map is non-uniform due to different pointing directions and roll angles with a relative minimum at the center of the field of view. This minimum occurs because only Field 1 and ObsID 1730 cover the central part of \m; the other observations either omit this area or the area around the nucleus was later masked out (see below).

The merging process combines source data from different off-axis angles for which the point spread function (PSF), response, and effective area may all be significantly different. It is therefore possible that the sensitivity and angular resolution for such sources is significantly degraded and requires that the source detection and characterization be done not only on the merged data, but on the individual fields and epochs as well.  For extracting the final source catalog only data from the ACIS-I chips I0-I3 were included because the best imaging performance was essential to achieve the highest sensitivity to point sources.  The S-array CCDs (S2 \& S3) were sufficiently far off-axis to suffer from reduced sensitivity and severe source crowding.

\subsection{Data Reduction}
Data reduction was carried out with CIAO v4.0.1 and CALDB v3.4.2. The first step in processing the data was the visual inspection, identification, and rejection of spurious features such as hot pixels and bad columns in the level=1 eventlists and bias maps. None of the \m\ data sets showed any of the rare large-scale distortions of the bias maps. We decided to generate our own bad pixel files as the standard bad column screening applied by the pipeline to the level=1 data products appeared to be too aggressive. In contrast to the diffuse X-ray emission, point sources typically have much smaller extraction regions and higher signal-to-noise (S/N) ratios. Therefore, a less conservative bad column screening with a slightly higher background should not seriously affect the source statistics, but would restore a significant fraction of the chip for source detection. We considered all warm columns which contributed more than 10\% to the total counts within a column of a source's extraction region for a given epoch to be 'bad', i.e., a threshold of 85 cts col$^{-1}$ 100\,ks$^{-1}$ for the front illuminated chips (I-array) had to be exceeded. The observation-specific bad pixel files were updated with the new bad column lists and as a result 3\%\,--\,15\% of the chip area which was previously contained in "warm" columns could be put back into the eventlists, including the neighboring columns next to the ones previously flagged as 'bad'. 

We next ran {\tt acis\_process\_events}, filtered on the background flags for the {\tt VFAINT} mode to improve the background rejection efficiency, corrected for bad pixels/columns, charge transfer inefficiency (CTI), and time-dependent gain variations, and removed the pixel randomization. The eventlist was then filtered for grades=0,2,3,4,6, status=0, and pipeline-provided good time intervals. Next, we checked for background flares by creating a background light curve from the source-free eventlists of the ACIS-I3 chip. 
We applied an iterative sigma-clipping algorithm ({\tt lc\_clean}\footnote{see http://cxc.harvard.edu/ciao/ahelp/lc\_clean.py.html}) to remove time intervals with count rates more than $3\sigma$ from the mean of each iteration, until all count rates were less than $3\sigma$ above the mean. The resulting effective exposure times are listed in the last column of Table~\ref{obs}. The same steps were also applied to ObsID 2023, except that {\tt check\_vf\_pha} was set to 'no'. 
All ObsIDs were checked for known processing offsets in the aspect solution using the aspect corrector\footnote{http://cxc.harvard.edu/ciao/threads/arcsec\_correction/index.html\#aspcalc} and any offsets found were corrected.

\begin{figure}[t]
\centering
\includegraphics[width=12cm,height=12cm,clip]{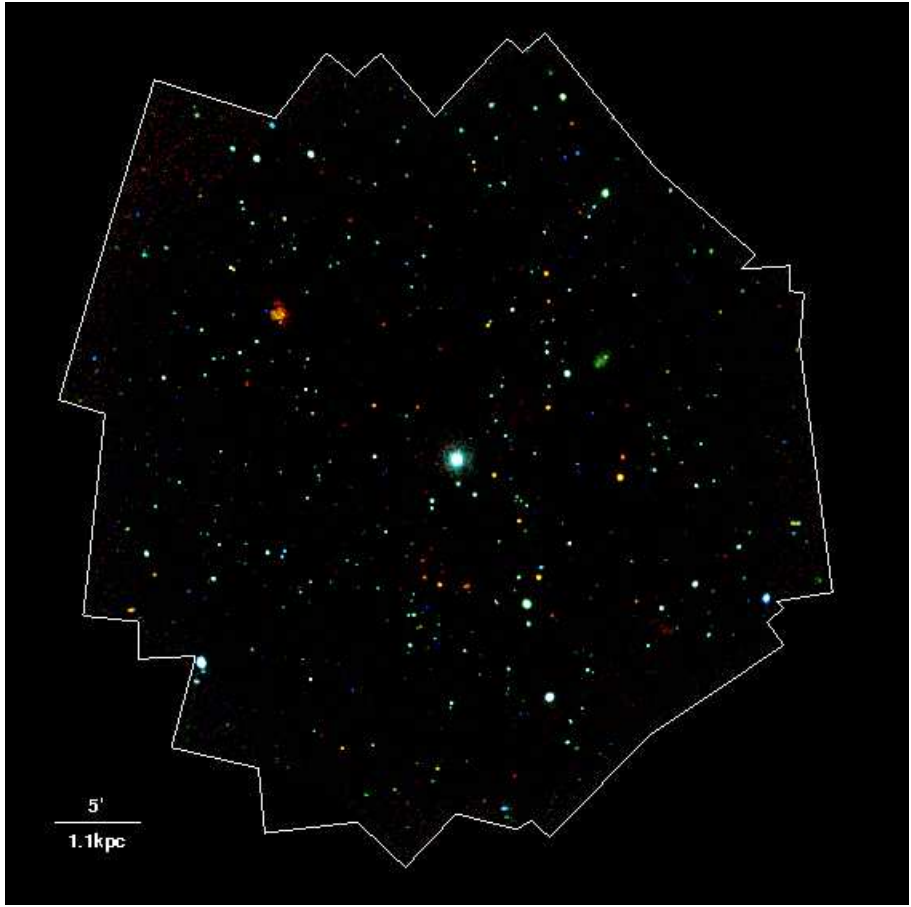}
\caption{\label{f2}
Multicolor exposure-corrected X-ray mosaic constructed from the full survey data. Red represents soft X-ray emission in the 0.35\,--\,1.0\,keV energy band, green represents medium hard X-rays (1.0\,--2.0\,keV), and blue denotes hard X-rays in the range of 2.0\,--\,8.0\,keV.}
\end{figure}
Transfer streaks from bright sources were removed from the individual level=2 event lists by masking out appropriately-sized regions. If a source was affected by the transfer streak in one of the observations, the analysis for this source was carried out without that observation. The resulting event lists were merged for each epoch (e1, e2) and field (e1+e2, see Table~\ref{obs}) and images were created in the following energy bands (keV): 0.35-8.0, 0.35-1.0, 1.0-2.0, 2.0-8.0 with bin sizes of 1, 2, 3, and 4. To create exposure-corrected images, appropriate exposure maps were constructed using {\tt mkinstmap} and {\tt mkexpmap} in CIAO using the same mask regions as for the corresponding event lists. For the instrument maps, spectral weights were computed assuming a power-law spectrum with $\Gamma=1.9$ that is typical of XRBs and AGN. An average \NH\ of $1\times 10^{21}$\,cm$^{-2}$ was assumed for the total line of sight \hi\ column density \citep{newton1980}, while the absorption in \m\ was determined from a weighted 21\,cm \hi\ map (D. Thilker, priv. comm.) to be $\sim$$0.4\times10^{21}$\,cm$^{-2}$. 

The composite X-ray image of \m\ is shown in Fig.~\ref{f2}. Soft X-rays (0.35\,--\,1.0\,keV) are displayed in red, medium hard X-ray emission (1.0\,--\,2.0\,keV) is shown in green, and hard X-rays (2.0\,--\,8.0\,keV) are represented in blue. In addition to a large number of point sources, the most prominent feature in this image is the nucleus of M33, M33 X-8, a stellar black hole candidate which is the brightest steady ultra luminous X-ray source in the Local Group \citep[e.g.,][]{dubus04,wewa09}. The extended reddish source to the NE is NGC\,604, the second largest \hii\ region in the Local Group \citep{tull08} and the elongated greenish emission to the NW of the nucleus is IC\,131, an enigmatic giant \hii\ region with an unusually hard X-ray spectrum \citep{tull09}.  

\subsection{Source Catalog Creation}
The general idea behind our source detection strategy is to use \wa\ \citep{freeman02} to determine the positions of potential source candidates and to input these positions to {\tt ACIS Extract} \citep[][\ace\ hereafter]{Broos02}, which is designed to determine the actual significance of a point source, allowing for variations in exposure time and PSF in different exposures.  Therefore, we deliberately created an initial source list that included many more sources than we expected to be statistically significant.
We applied an iterative procedure in which the source candidate list was filtered with increasing source significance thresholds and the output was checked for potentially lost sources. The screening process continued until only a small number of spurious sources was present in the final source list (see below for details).
\subsubsection{Point sources}\label{subsec-ps}
We used the CIAO tool \wa\ \citep{freeman02} to identify candidate sources in an image by correlating it with a "Mexican Hat" wavelet function using different scale sizes to account for sources located at different off-axis angles. For each scale size, the input image is correlated with the wavelet function and pixels with a sufficiently high correlation coefficient are removed as 'source' pixels (the removal of source pixels from the image is called "cleansing"). 
In order to decide on reasonable values for the source significance ($sigthresh$) and cleansing thresholds ($bkgsigthresh$), the maximum number of cleansing iterations per scale ($maxiter$), and spatial scale sizes ($scale$), we used fields 3 and 4 \citep[epochs 1, 2, and 1+2, see][for their locations]{ppp08} as testbeds for a series of \wa\ runs in which we evaluated several combinations of these parameters. Based on these tests, we decided to run \wa\ with $sigthresh$\,=\,$10^{-4}$, to include a substantial number of unreal sources ($\sim$\,100 per 1\,Mpix) but also to catch all real sources, $bkgsigthresh$\,=\,$10^{-2}$ \citep[see][]{freeman02}, and $maxiter=5$, to avoid background contamination by additional source counts. All other parameters were left at their default values. If we, for example, would have chosen values for $sigthresh\le 10^{-6}$, we would have lost a handful of potentially real sources \citep[see also][]{nandra05}, which supports our decision to not let \wa\ do the source culling. 
We also tested to what extent different binnings affect the number of sources found by \wa. For this purpose it is important to keep in mind that the wavelet 'scales' match the pixel size for each binning. If we, e.g., wanted to search for sources with a radius of 4 pixels, we used $scales=4$ and $bin=1$, and compared the number of potential sources with the one for $scales=1$ and $bin=4$. We found that the 'bin=1(8), scale=64(8)' lists did not contain any additional statistically significant sources that were not already included in the lists with lower bin/scale sizes. However, up to 3 source candidates with 9\,--\,31 counts were detected in the remaining lists with higher bin sizes that were not in the lower binned data. Therefore, we chose to run \wa\ on images with bin sizes of 1, 2, 3, and 4 and $scale$ sizes that ranged from 1 to 32 pixels (in a power-of-2 sequence) in each of the subbands. With these parameters we are also able to detect extended sources on spatial scales of up to $\sim$1\arcmin\ which are needed to include giant \hii\ regions like NGC\,604 or IC\,131 (see Sect.~\ref{subsec-es}).

As can be seen from the multicolor composite image shown in Fig.\ref{f2}, some sources show rather soft X-ray emission which is typical for SNRs, \hii\ regions or supersoft sources. Others emit much harder X-rays as one would expect from AGN or X-ray binaries. Therefore, we ran the source detection in several energy bands that are tailored to the different kinds of sources. We adopted the following energy bands (in keV): 0.35-8.0, 0.35-1.0, 1.0-2.0, 2.0-8.0. 
Finally, we ran \wa\ on all \chase\ fields on each individual epoch (e1, e2, and e1+2) in four different energy bands and binnings as well as on the two archival fields (ObsIDs 1730, 2023), resulting in a total of 368 source candidate lists. Because many of the sources in the various lists are duplicates of one another and to save substantial processing time later in \ace, the different \wa\ source candidate lists were merged sequentially according to the following scheme:
1) Match the sources among different binnings of the same field, epoch, and energy, choosing for each source the position belonging to the detection at the smallest binning (this method assures the best centroiding of the source and the matching creates a single list for each energy of each epoch of each field).
2) Match the sources among different energies for the same field and epoch, choosing for each source the position belonging to the detection with the highest significance (detections with the highest significances tend to have the higher count rates with respect to the background, and thus better centroiding; a single list for each epoch of each field is created).
3) Match the sources among different epochs of the same field, choosing for each source the position belonging to the detection with the highest significance (this matching creates a single list for each field).
4) Match the sources among different fields/archival ObsIDs, choosing for each source the position belonging to the detection with the smallest PSF (this matching creates a single list for the entire galaxy).

PSF sizes and source significances were taken from \wa\ output (parameters $psf\_size$ and $src\_significance$). The resulting merged \wa\ source list contained positions of about 9700 source candidates and was used as the input to \ace.

\ace\ is a multi-purpose source extraction and characterization tool which determines source and background count rates and calculates fluxes and source significances for all source candidates within a number of user-specified energy bands. Among other tasks, it also extracts and groups spectra, generates the appropriate response matrices, and constructs lightcurves and checks for time variability \citep[see][for details]{Broos02,Broos10}.
An important feature of \ace\ is its ability to produce source and background extraction regions by simulating the \chandra\ PSF at a given source position, i.e., off-axis and azimuthal angle. This capability is essential for sources in overlapping fields as for each individual observation a different extraction region is required in order to optimize the S/N for that source. Moreover, \ace\ can shrink the source extraction regions so that they do not overlap. This is important when source crowding is an issue, as for example for the circumnuclear region or NGC\,604. 

For the \chase\ First Look Catalog \citep[][FLC hereafter]{ppp08} we adopted a traditional S/N ratio estimate ($src\_significance$) to distinguish between real and false source detections. Because this criterion seemed to be too conservative (a $3\sigma$ source had to have 15 counts in the case of infinitesimally low background and therefore biases the sample towards higher count rates), we have chosen the Poisson probability of not being a source ($prob\_no\_source$, $pns$ hereafter) provided by \ace\ as our new threshold criterion, which also takes the uncertainty of the local background estimate into account. $pns$ represents the Poisson probability that all of the counts within the source extraction region are actually from the background, i.e., source candidates with low probabilities are most likely real sources and vice versa.
Source pruning was done by filtering the source candidate list iteratively with decreasing $pns$ thresholds until a suitably low false source detection rate was reached. This false source detection rate was determined by a number of \wa\ and \ace\ test runs on the eventlist for field 4 (epoch 2) from which all sources were removed and for which pixel coordinates were randomized, to smooth out possible residuals from the source removal and to get a uniform background. 
We determined that $pns$ values around $\simeq$10$^{-6}$ provided a suitably low false detection rate in agreement with the work of \citet{nandra05} and \citet{george08}.  We decided to adopt their value of $pns=4\times10^{-6}$ which results in $\sim$0.5 false detections per 1\,Mpix.

As the source extraction and characterization in \ace\ works only on eventlists we had to merge the aspect solution files for each field and epoch using the CIAO tools {\tt dmmerge} and {\tt dmsort}. Due to scheduling constraints, ObsID 7208, which is part of Field 6 epoch 1, has a significantly different roll angle and much lower exposure, which is why we decided to analyze this observation separately. The merging process substantially reduced the number of AE runs from 27 to 17.
\ace\ was run first on each individual field and epoch (including ObsIDs 1730, 2023, and 7208) using the full list of $\sim$9700 sources and with no $pns$ filter applied in order to get a first estimate of the $pns$ value for each source candidate. Source and background extraction regions were created with the {\tt ae\_make\_catalog} recipe, using 90\% of the PSF at a primary energy of 1.49\,keV ($psf\_frac=0.9$) and setting $mask\_fraction$ and $mask\_multiplier$ to 0.95, and 1.2, respectively. The latter two parameters control the size of the inner radius of the background region (see \citet{Broos02} for details). The outer radius of the background annulus was allowed to expand until a minimum of 50 background counts in each background region was achieved ($min\_num\_cts=50$).  Next, {\tt ae\_standard\_extraction} was executed which extracts the source eventlists and source and background spectra. 
Source properties, such as the $pns$ and flux values, were computed in the 0.5\,--\,8.0, 0.5\,--\,2.0, 2.0\,--\,8.0, 0.35\,--\,8.0, 0.35\,--\,1.1, 1.1\,--\,2.6, 2.6\,--\,8.0, and 0.35\,--\,2.0\,keV energy bands. 
These bands were adopted for the FLC \citep{ppp08} and the \chase\ SNR catalog \citep[L10 hereafter]{long10}; some of these bands were also used by other major X-ray surveys, such as ChaMP \citep{kim07} and the CDF-N \citep{alex03}.
After that we applied the {\tt /merge\_observations} stage which generated a composite PSF for each source candidate from all observations, a merged spectrum, a merged auxiliary response file (ARF), a merged response matrix file (RMF), and a merged source eventlist. In the final step the {\tt /collate\_filename} stage was run which combined all source parameters and properties into a single FITS table (hereafter simply called output).

Due to the liberal choice of $sigthres$ in \wa, the majority of sources in the full initial list are unreal and source crowding poses a serious issue. The vast number of sources seriously reduces the available background area and because of the large-scale diffuse X-ray emission in \m, the average background could be artificially raised and result in a systematic underestimate of the $pns$ value for every source. Therefore, the output from the first \ace\ iteration was filtered with $pns=10^{-2}$, a value of which we were certain that no real sources will be lost, and \ace\ was run a second time to restore the background area and to get revised source properties. This filtered source list contained 1800 sources which is still a factor of $\sim$5 higher than the total number of sources in the FLC and further filtering was applied as described below.

We overlaid the region files for each field generated by \ace\ onto the corresponding eventlists and we checked if any obvious sources were missed or displaced as a consequence of the merging of the \wa\ source lists. It turned out that 8 sources were lost/displaced and all, except one, were detected by \wa, implying that they were lost during the merging process. The source candidate which was not detected by \wa\ was a special case, as it is a faint source more than 8\arcmin\ off-axis which is located between two very bright X-ray sources. The missed source was accounted for in the next \ace\ run by adopting appropriate center coordinates. The displaced sources, however, raised doubts whether the adopted merging scheme for the \wa\ sourcelists was the correct one. Instead of using the position of the most significant source for merging the different bin sizes, we alternatively used the position of the lowest binning. Although this solved the issue with the missed sources, there were new cases for which the previous combination was more appropriate. We therefore decided to modify the positions of the source candidates by hand, which resulted in the repositioning of 36 relatively faint sources.

We also created thumbnail catalogs for all observations of each source and its neighbors and overplotted source and background regions together with the estimated $pns$ value, net counts, and off-axis angle. By visually inspecting these catalogs, we deleted obviously misidentified source candidates, e.g., multiply identified sources or those for which the PSF contours were systematically offset from the obvious source position in one observation and which were not present in other observations. Both issues are residuals from the merging process where \wa\ source lists for different binnings and sources at different off-axis angles were combined.  
Even with a reduced number of 1800 source candidates, a handful of cases existed in which source crowding was an issue. This typically occurred in the off-axis data in which the PSF was larger than in the on-axis data. If a less-significant source was located close enough to a more significant source such that the extraction region of the more significant source was decreased to less than 66\% of the PSF size, the less significant source was removed for that observation. The on-axis data for the less significant source were retained for the characterization of the source.
In case a source looked dubious but was isolated and its extraction region did not affect neighboring sources, it was kept. 
The screening process removed a total of 154 sources from the \ace\ output list and was done by three team members individually in order to compensate for subjective biases. This approach was always used when source candidates had to be added, removed, or repositioned.

The nucleus, because of heavy pile-up, was treated specially. We included only the on-axis pointings for imaging, and only the off-axis pointings for the spectral analysis. In the on-axis Field 1 observations as well as in ObsID 1730 pile-up is most pronounced and the source has a highly peculiar ``Mexican hat''-like shape. As a consequence, \wa\ reported 31 source candidates within a radius of 24\arcsec, which almost all looked spurious. We therefore replaced all sources within that region with a single source region centered on the position of the nucleus and carried out a separate AE run on F1e1, F1e2, and ObsID 1730 with a $pns$ value of $10^{-2}$. Because of the pile-up and a heavily distorted PSF in the on-axis pointings, we cannot tell whether any of these replaced sources are real. For the separate AE run, 30 sources within 1.7' from the nucleus were considered and those which did not pass the $pns$ threshold were removed from the source directory tree. This step fortunately removed the source crowding around the nucleus in the off-axis pointings (F4e2 and F6e1). As pile-up is less pronounced in these observations, the spectroscopic analysis of the nucleus was only carried out with these two off-axis pointings.

Next the full \ace\ output was filtered with a $pns=10^{-3}$ to further reduce source crowding and to remove other spurious sources, yielding 918 source candidates. Sources which did not pass the $pns=10^{-3}$ filter were visually examined to ensure that potentially real sources (e.g., transient sources) were not erroneously removed. This number had to be trimmed down further as the source list still contained a substantial number of spurious source candidates. 
For the source screening process, the following nine energy bands were considered (in keV): 0.5\,--\,8.0, 0.5\,--\,2.0, 2.0\,--\,8.0, 0.5\,--\,1.0, 1.0\,--\,2.0, 2.0\,--\,4.0, 4.0\,--\,8.0, 0.35\,--\,1.0, and 0.35\,--\,8.0. We decided to include a source if [1] the $pns$ value in any of the nine energy bands is $\le 10^{-5}$ for the combined data set or [2] the $pns$ of a source which was rejected by [1] is $\le 10^{-5}$ in any individual ObsID and in any of the nine energy bands. Criterion [2] should catch potential transient sources and by including the energy bands mentioned before, we account for different types of sources with different spectral energy distributions. If we would have considered merely the 0.35\,--\,8.0\,keV band, $\sim$3\% of significant sources would have been lost. A sanity check was carried out to ensure that no potentially real sources were missed by examining again those source candidates which did not pass the $pns$ threshold. This latest screened list contained 721 source candidates. Further trimming was necessary as the visual inspection of the catalogs still indicated a significant number of false sources. We applied the same two $pns$ criteria as mentioned above and ran \ace\ with the final $pns$ value of $4\times 10^{-6}$. 
The analysis of the thumbnail images of those sources in the catalog and those who did not pass the final $pns$ filter indicated, that no obvious source candidates were lost and that source crowding was no longer an issue. However, three more corrective actions were taken. 
First, there were sources which were located at or which extended beyond the chip edges, so that only a fraction of the source photons was collected. The extraction regions of these far off-axis sources were heavily distorted and would have caused incorrect $pns$ and flux values due to zero or strongly reduced exposure. Therefore, those observations were removed from the source directory tree, provided there was more than one observation left for this source.
Second, sources on transfer streaks had to be dealt with. There were 9 sources covering, at least partly, the masked-out region of the transfer streaks. By narrowing down the width of the mask region, 4 sources could be fully restored and for the remaining 5 sources, we had to remove the corresponding observations from the source candidate list. 
Third, new source and background extraction regions for the extended sources in \m\ had to be created in order to obtain proper estimates of the source properties (see Sect.~\ref{subsec-es} for details).
With these modifications applied, AE was run a second time with a $pns$ value of $4\times 10^{-6}$, but this time enabling the {\tt /fit\_spectra} and {\tt /timing} stages in AE.

As this source list has been tailored such that each source has proper extraction and background regions, we can now examine how the source properties, such as $flux2$\footnote{$flux2$ computes the flux by assuming a mean ARF in the band of interest and dividing the net counts in that band by the mean ARF (see the \ace\ manual for details)} and $net\_cnts$, change if the number of background counts within the background extraction region increases from 50 to 100. This was tested by running AE on the final source list and setting $bkg\_cnts$ to 100 and comparing the results with the corresponding 50 counts source list. One general trend emerged, namely, sources with 100 counts in the background extraction region tended to have slightly higher source net counts and fluxes than the ones with 50 background counts. This could mean that the background is higher closer to the sources due to diffuse emission in \m. As the difference between the two AE test runs is $<10\%$ and therefore less than the uncertainty in the flux calibration, we continued to work with $bkg\_cnts=50$.

Now that source crowding issues have been solved by iterative pruning of the source list and because the spatial resolution of \chandra\ will be unrivaled in the coming decades, we decided to spend extra time on improving source positions. For this purpose we ran the {\tt /check\_positions} stage of \ace\ which provides three additional position estimates. Besides simply adopting the position from \wa\ (method catalog), a mean peak position is calculated by centroiding on the merged data (method mean), a correlation between the merged image and the merged PSF is carried out (method corr), and the merged image is reconstructed using a Maximum Likelihood algorithm and finding the peak in the reconstructed image \citep[method ML, see][for details]{Broos02}. If the individual observations for a given source suggested different position estimates, the observation for which the source was closest to on-axis and/or had the lowest $pns$ was chosen to determine the position of the source. In case the estimates were very close to each other, the original position from \wa\ was kept and if none of the estimates was preferred, an appropriate position of the source's center was determined by eye. 
In order to provide consistent results between L10 and this work, we adopted the same positions and extraction regions used by L10. The source repositioning required rerunning {\tt ae\_make\_catalog}, {\tt ae\_standard\_extraction} as well as the {\tt /fit\_spectra}, {\tt /merge\_observations}, and {\tt /collated\_filename} stages of \ace. 

The positional uncertainties provided by \ace\ were sometimes unreasonably small ($\sim$0.01\arcsec), since systematic errors such as aspect reconstruction are not considered. To achieve a more accurate positional uncertainty, we applied the following strategy: 1) If there is only one observation for a given source, the distances between the adopted catalog position and those suggested by the other three methods (mean, corr, ML) are calculated. The positional uncertainty is then simply the average distance to the catalog position. 2) If there are two or more observations for a given source, all observations are used to calculate the average distance for each position estimate with respect to the catalog position. The most discrepant position estimate is discarded and the average positional uncertainty is determined from the two remaining position estimates. 3) If the uncertainty is less than 0.5\arcsec, the positional uncertainty was set to 0.5\arcsec. This value is taken from the CXC absolute astrometric accuracy web page\footnote{http://cxc.harvard.edu/cal/ASPECT/celmon/} and also accounts for a possible systematic shift which might have been introduced by the merging of the different fields. To see how our astrometric accuracy compares to other X-ray catalogs, such as the ChaMP survey \citep{kim07}, we computed a histogram using the values for the positional uncertainties listed in column (5) of Table~\ref{sourcelist} and compared it to that provided by \citep{kim07}. Both histograms are in good agreement; their distribution peaks at 0.7\arcsec, ours at 0.6\arcsec\ and also shows outliers well above 1\arcsec. Our largest positional error of 3.5\arcsec\ is for a faint source (\#179) that is far off-axis.

\subsubsection{Extended sources}\label{subsec-es}
Although all sources in the catalog presented here were detected in our point source search, some of them are actually extended. Our approach to determining this was as follows:
we used the energy-filtered eventlists (0.35-2keV), subtracted the point sources, created (if necessary) custom-made ellipse regions to include the remaining counts for those sources which showed obvious excess counts, and calculated the number of source counts from the polygon region ($N_p$) provided by \ace\ and the total number of counts within the ellipse region ($N_e$). 

A necessary condition for a source to be extended is that $\zeta=1 - (N_p/N_e) > (1 - psf\_frac$) in epoch 1 or 2 ($psf\_frac=0.9$). However, for the source to be 'confidently' or 'possibly' extended, we calculate the $3\sigma$ uncertainty and include a term which accounts for the uncertainty within the PSF model (here set to 1\%). Therefore, if $\zeta > 0.1 + 3 \sqrt{d\zeta^2 + 0.01^2}$, with $d\zeta = \zeta \sqrt{1/N_p+1/N_e}$, then the source is 'confidently' extended, else if $\zeta \le 0.1 + 3\sqrt{d\zeta^2 + 0.01^2}$, the source is 'possibly' extended. If $\zeta\le0.1$ the source is not extended.

Most of the extended source candidates were observed multiple times, which means that they can have different extension flags. In the extreme case in one observation a source can be confidently extended, in another it could be not extended. We applied the following merging scheme if the sources had different flags in different observations:
if the source was confidently/possibly extended (or vice versa), the source was flagged as possibly extended. If the source was not extended/possibly extended (or vice versa), the source was flagged as possibly extended. 
For all possibly and confidently extended sources in the FLC or in the SNR study (L10), we estimated an intrinsic source size using an elliptical region. To account for the varying PSF with off-axis angle and roll, the ellipse was ``puffed out'' using the appropriate AE-generated PSF polygon region (see L10 for more details on this procedure). These regions, together with new source and background eventlists and exposure maps replaced the original point source data. For consistency, we adopted the Long et al. regions for extended sources when available.
In Table~\ref{crossref} (see Appendix \ref{app}) the column labeled ``Extended'' flags each source either as not extended (=0), extended (=1), or possibly extended (=2). Among the 70 extended source candidates are the 23 extended SNRs listed in L10. L10 detected 82 (58) SNRs at 2$\sigma$ (3$\sigma$). We did not detect all of them due to our initial requirement, that the source had to be detected as a point source. 

It should be noted that the $3\sigma$ criterion plus a relatively low uncertainty in the PSF model ($<$5\%) provides the best source classification compared to 3$\sigma$ and 5$\sigma$ criteria which neglect uncertainties in the PSF model. A 5$\sigma$-screening especially appears to be problematic as clearly extended objects would have been downgraded from confidently to possibly extended.

\section{Results and Discussion}\label{sec-resdis}
\subsection{The Source Catalog}\label{sec-srccat}
The final \chase\ source catalog contains 662 sources\footnote{The source catalog is also available in FITS format at: http://hea-www.harvard.edu/vlp\_m33\_public/}. Their positions and properties, such as the source significances ($pns$ values), net counts, and photon fluxes (in eight different energy bands) are listed in Tables~\ref{sourcelist}, \ref{mergedpns}, \ref{netcts}, and \ref{mergedphotfl}. We will use this information to carry out a detailed spectral analysis of the 15 brightest X-ray sources (excluding the SNRs studied by L10 and X-7, the eclipsing XRB studied by \citet{pietsch06}), to create hardness ratio diagrams to learn about the different source populations in \m\ (SNRs, XRBs, etc.), and to create an X-ray LF.  For the latter we first have to cross-reference the \chase\ sources with the other catalogs and observations from other wavelengths to screen out sources which are definitely not associated with \m\ and to identify sources which may be in \m. Finally, we search for time-variable sources and supersoft X-ray sources.

\subsection{Comparisons with other Catalogs and other Wavelengths}\label{sec-comp}
We cross-correlated the final \chase\ source catalog with the FLC, the \xmmnewton\ catalogs of \m\ from PMH04 and MPH06, the \chandra\ catalog from G05, the SNR catalog from L10, the 2MASS All-Sky Catalog of Point Sources \citep{cutri03}, and the USNO-B1.0 catalog \citep{monet03}. 
We also utilized optical follow-up spectroscopy of 116 X-ray sources obtained with the Hectospec spectrograph \citep{fabric08} attached to the MMT to search for and identify possible counterparts. 
Because there is a significant number of X-ray sources that are neither listed in the above catalogs, nor covered by our spectroscopic follow-up observations, we also used multi-wavelength imaging data to attempt to determine the nature of each source.

\subsubsection{Comparison to other X-ray catalogs}\label{sec-xcat}
Although the final \chase\ X-ray catalog and the earlier FLC are based on many of the same observations, the reduction and analysis techniques are very different. Of the 394 sources in the FLC, only one is missing from the current catalog, source FLC 191, which is in the highly confused nuclear region which we replaced by a single extraction region (cf. Sect.~\ref{subsec-ps}). Some of the FLC sources have substantial offsets from the current catalog position as those sources were detected in FLC only at large off-axis angles. As a result, the positions reported here are to be preferred from those in the FLC. 

There has been a substantial effort to observe the entire disk of \m\ with a somewhat shallower \xmmnewton\ survey by PMH04 and MPH06. Of the 233 sources listed by PMH04 which fall within the \chase\ FOV, 217 appear in our catalog. Of those that remain, 7 appear coincident with diffuse emission (PHM-89, 101, 125=IC\,131, 170, 266, 299=NGC\,604, and 365) but were detected as a single source. Three others are in regions with low \chandra\ exposure (280, 281, and 352).
Of the remaining 6 PHM04 sources that ought to have been detected (52, 223, 245, 313, 328, and 332), all but one (PMH-245) have {\it XMM} fluxes near the detection limit of the PMH04 survey ($10^{-15}$\,erg\,s$^{-1}$\,cm$^{-2}$) and may well be false detections. It appears likely that PMH-245 is a transient source.

Of the 39 sources in MPH06 that are not in PMH04, 26 fall in the \chase\ FOV. Of these, three match \chase\ sources. Two additional sources were associated with bright \hii\ regions (one inside NGC604 and one in NGC595) where source crowding and confusion with the large-scale diffuse X-ray emission is an issue. Another four sources fall in regions with strong diffuse emission. None of the 17 remaining sources have obvious counterparts, diffuse or otherwise in the \chase\ data.
Since the MPH06 catalog was constructed only from individual observations rather than stacked data, the detection limit is slightly higher. Of the 17 unmatched sources, about half are fainter than $2\times10^{-15}$\,erg\,s$^{-1}$\,cm$^{-2}$, near the detection limit. The other half should have been detected in our survey, and hence are likely transient or variable sources. 

We detect 211 of the 261 sources cataloged by G05 in their analysis of early \chandra\ observations of \m. One source is covered by the S2 chip and therefore outside the \chase\ FOV. The remaining 49 source candidates have no obvious \chase\ counterparts, neither in the individual event lists, nor in the merged ones. 
As already discussed by \citet{ppp08}, the main reason for this discrepancy is clearly the liberal source selection criteria chosen by G05 (they demanded only a detection by \wa\ at any significance level). Forty four of these 49 sources were filtered out during our iterative source pruning process while 5 sources were not even considered to be candidates in our \wa\ runs. It is therefore possible that most, if not all of the 49 sources listed in the G05 catalog are actually spurious sources. 
Moreover, the reason why the G05 survey, which is much shallower than \chase, apparently reaches a similar limiting sensitivity is that G05 used source significances provided by \wa. The source with the lowest number of net counts in G05 is CXO\,J013400.7+304138, which has 1.7 net counts in the 2.0\,--\,8.0\,keV band. The source was not detected by G05, neither in their soft band (0.3\,--\,2.0 keV) nor in their broad band (0.3\,--\,8.0 keV). There are other sources in the G05 catalog that have for example 2.8 and 2.9 net counts.  Such sources did not pass our source significance threshold. Our results of the cross-correlation are summarized in Table~\ref{crossref}.

\subsubsection{Comparison with optical follow-up spectroscopy and other multi-wavelength data}
The 662 sources represent a wide variety of objects, ranging from foreground stars to XRBs, SNRs, and \hii\ regions in \m, to background AGN. Most of the sources are faint, and therefore, to make progress in understanding the nature of the sources, we have begun an optical spectroscopy study of 116 of the 662 \chase\ sources using Hectospec, a fiber-fed spectrograph attached to the MMT. The fibers were positioned such, that, if there was an optical counterpart to the X-ray source, the fiber was put at the position of the optical source. If no optical emission was present the fiber was put down at the position of the X-ray source. The spectra cover the wavelength range from 3700\AA\ to 9150\AA\ and were obtained with a fiber diameter of 1\farcs5, which provides a resolution of 6.2\AA. This resolution is sufficient to classify all of the spectroscopically observed sources. Standard data reduction techniques were applied and spectrophotometric standard stars were used to flux calibrate the spectra and to correct for the wavelength-dependent instrument response. 

Since we do not yet have spectra of the majority of sources, we have also attempted to gain some idea of the nature of the sources by examining 
additional multi-wavelength data including archival imagery from \spitzer\ (3.6, 4.5, 5.8, 8.0$\mu$m), 2MASS (J, H, K), CFHT MegaCam (u, g, r), HST (f435w, f555w, f814w), data from the Local Group Galaxies Survey \citep[LGGS,][]{mass06}, such as  U, B, V, R, I, H$\alpha$, [\oiii ], and [\sii ] images, as well as archival \galex\ (NUV, FUV) and \swift/UVOT (UVW2, UVM2, UVW1) data. 

The infrared data are sensitive to Milky Way stars, red giants, and associations of older stars in \m, galaxies, and background AGN. Similarly, the USNO catalog is sensitive to Milky Way stars, compact clusters in \m\ and AGN, with the further selection that the USNO survey excludes the central disk of \m.
 The high spatial resolution of the optical data, especially of the HST data, provides a sensitive tool to search for optical counterparts to the X-ray sources, while the UV emission is a good tracer of the young stellar population in \m\ and of AGN-type background sources.  


Combining the results from the spectral analysis and the imaging analysis, we divided the X-ray sources with counterparts into the following seven groups: (1) foreground stars (FSs), (2) stellar sources in \m\ (i.e. objects with star-like spectra, such as stellar associations), (3) QSOs/AGN (sources with broad and redshifted emission lines), (4) galaxies (i.e. obviously extended sources with redshifted emission lines, but no broadened emission line components as typically seen in QSOs/AGN), (5) XRBs (including known XRBs and XRB candidates from HP01, PMH04, MHP06, and W08), (6) SNRs (SNR-like spectra, to be distinguished from \hii\ regions by significantly stronger [\sii]\ emission lines), and (7) non-stellar sources (everything which is not classified as stellar, but looked extended in the LGGS/MCam/HST data and did not match any of the other classes).

\begin{figure}[!t]
\centering
\hspace{-0.35cm}
\includegraphics[width=15cm,clip]{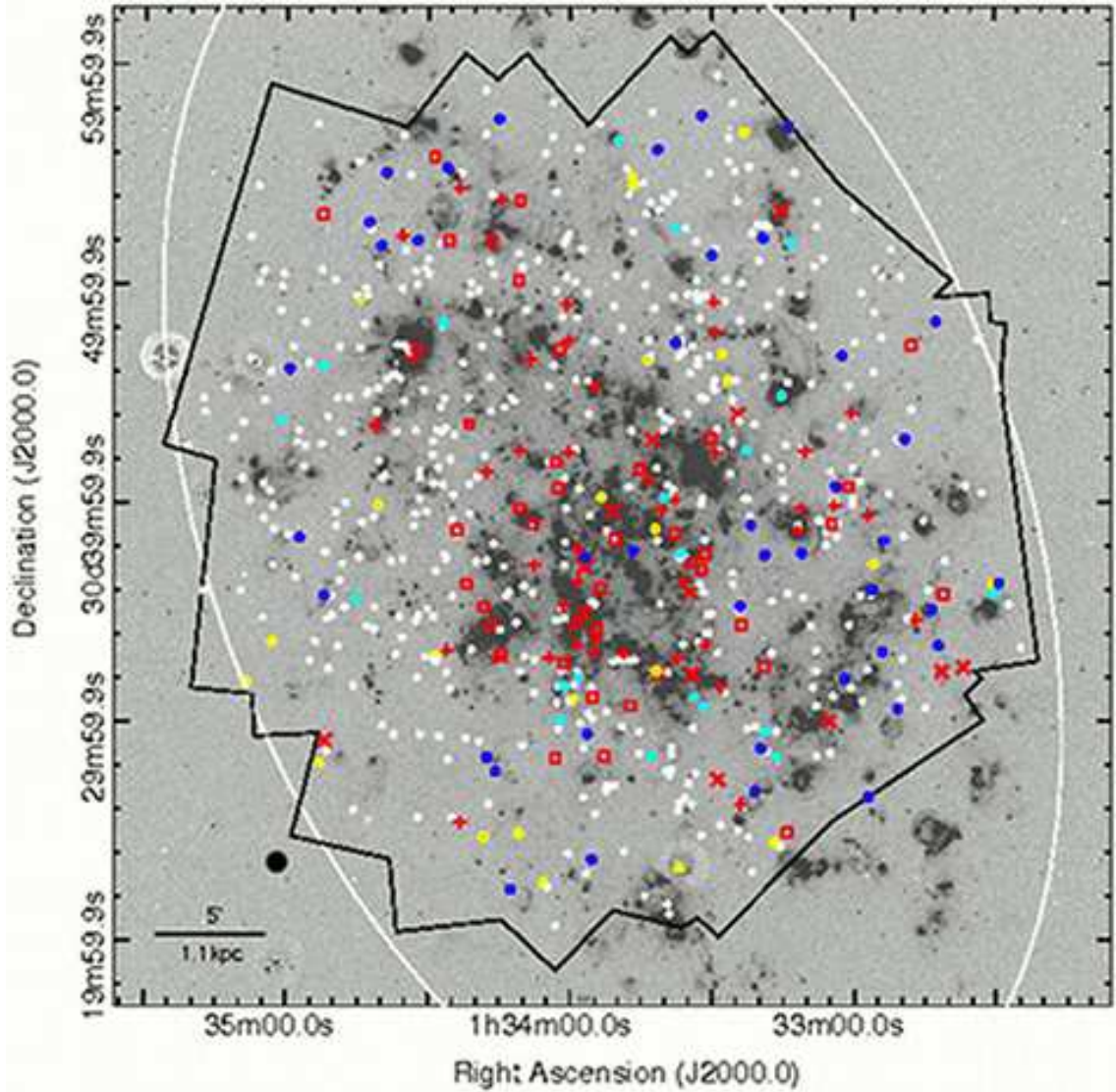}
\caption{\label{fclass} \chase\ sources overplotted on the same image as shown in the left panel of Fig.~\ref{f1}. Red sources are considered to be located in \m\ (crosses represent SNRs from L10, ``x'' symbols are known XRBs and XRB candidates, and squares are stellar sources). Blue circles stand for QSOs, AGN, and background galaxies, cyan circles are for non-stellar sources, yellow circles represent FSs, and unidentified sources are plotted in white.
}
\end{figure}

\begin{figure}[!ht]
\centering
\includegraphics[width=\textwidth,clip]{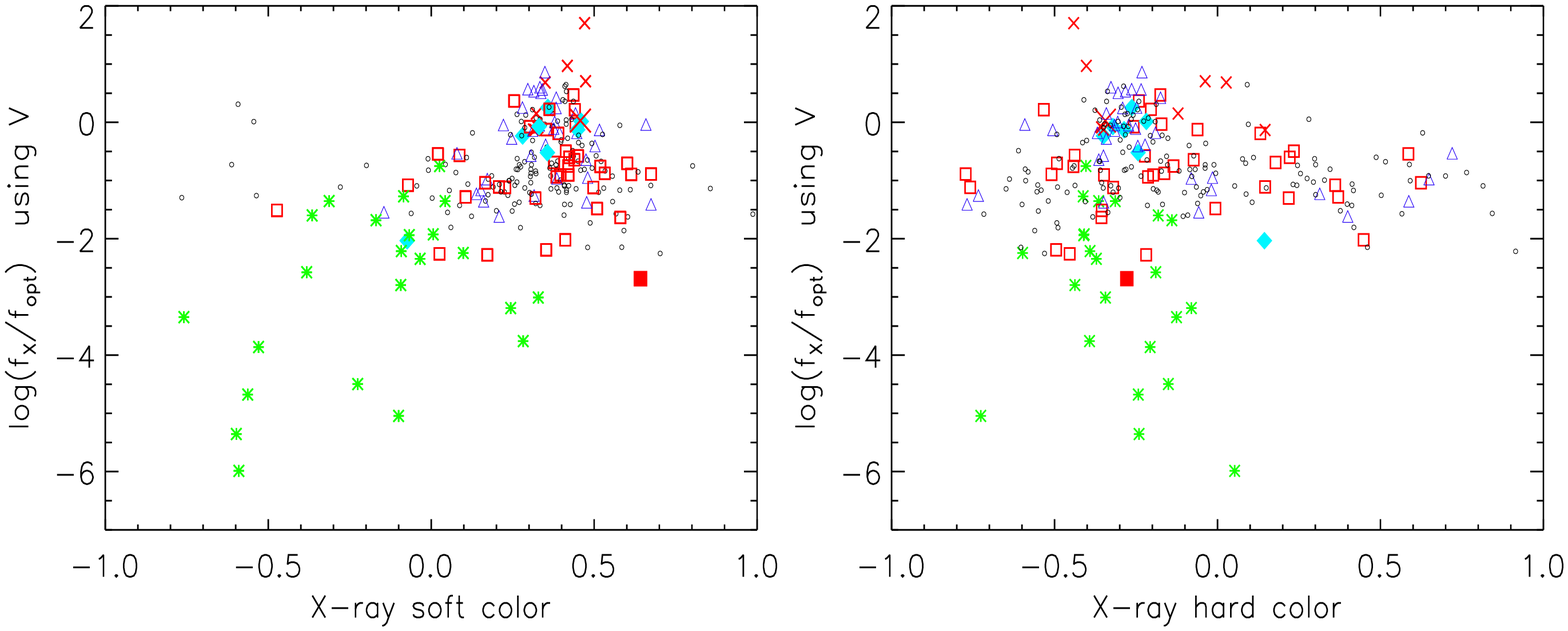}
\caption{\label{f5} X-ray (0.35\,--\,8\,keV) to optical (V band) flux ratios as a function of soft (M-S)/(S+M+H) and hard (H-M)/(S+M+H) X-ray color. A similar symbol/color scheme as in Fig.~\ref{fclass} was adopted, again using red symbols for sources which are considered to be in \m\ (the large red cross represents the nucleus of \m, while the solid red square denotes one of the two globular clusters). Green asterisks are FSs, blue triangles represent galaxies, QSO, and AGN, cyan diamonds are non-stellar sources, and black sources have no suggested identifications.
The sources plotted here are a subset of the X-ray catalog, as not every X-ray source has an optical counterpart in the LGGS data.}
\end{figure}

As a result of this effort, we have suggested classifications for 183 \chase\ sources. Among them are 23 foreground stars, 40 stellar sources (of which 2 are classified as globular clusters, GC hereafter), 20 QSOs and AGN, 21 galaxies, 14 XRBs, 45 SNRs (those from L10), and 20 non-stellar objects (see Table~\ref{crossref}, column ``Source type'' for details). Considering the uncertainty associated with the positional coincidence of a source at X-ray and at other wavelengths, we do not expect every source to be classified correctly. However, we are confident that the bulk of sources are identified correctly. Fig.~\ref{fclass} shows the spatial distribution of the afore mentioned source classes across the FOV. The distribution of SNRs nicely follows the spiral arms of \m, while the observed AGN candidates seem to be located in the outskirts of this galaxy. We have optical spectra for all stellar sources, except one GC source, and it is conceivable that there is a significant number of XRBs among the 39 sources whose spectra are indistinguishable from other stellar sources.

A clear separation between foreground stars and sources with stellar-like spectra (e.g., globular clusters or star-forming regions in \m) and background AGN can be accomplished by plotting X-ray to optical flux ratios vs. X-ray colors as shown in Fig.~\ref{f5}. These plots contain 229 of the 662 \chase\ sources and are those for which an optical counterpart could be identified.

We note that there may be a small systematic shift between the coordinate systems of the \chase\ catalog and the other surveys mentioned above. These shifts are below the typical \chase\ source positional error of 0\farcs5, and could arise from small (some 0\farcs1), systematic positional errors of each data set. Based on sources with well known optical counterparts in the LGGS, we find that the \chase\ positions are systematically offset by $-$0\farcs4 in RA and $-$0\farcs2 in Dec relative to the LGGS ones. We also cross-correlated \chase\ with the USNO-B1.0 and 2MASS All Sky point source catalogs (48 and 78 matches, respectively).  The mean offset between the \chase\ sources and the matched USNO (2MASS) sources was $\Delta$RA\,$=$\,0\farcs12 (0\farcs11) and  $\Delta$dec\,$=$\,$-$0\farcs06 ($-$0\farcs07), respectively. 

\subsubsection{Comparison with sources in globular clusters}
Given the long-standing interest in the X-ray globular cluster LF for spiral galaxies, we attempted to find the discrete X-ray sources associated with GCs by cross-correlating our source list with the (mostly non-overlapping) catalogs of GCs in \m\ from \citet[][SM07 hereafter]{sara07} and \citet[][ZKH08 hereafter]{zloc08}. For the ZKH08 catalog, which was created from the CFHT Megacam survey of \m\ and covers the outer part of our survey region, we selected only sources that were likely to be clusters (their type 1 and 2) and accepted as matches those X-ray sources that were located within 1 FWHM of the cluster center. The SM07 catalog, which is a heterogeneous collection of older catalogs and covers the bulk of our survey region, does not list a cluster size, so we used the mean cluster size from ZKH08 ($1\farcs8$). We selected all sources listed as cluster candidates in that catalog, and accepted as matches those X-ray sources within $1\farcs8$ of the cluster center, roughly the same criterion used for the ZKH08 catalog.
Between these two lists we found two potential matches. \chase\ source number 393 is well matched to the GC SMS275 (which was verified with the Megacam and HST data) and \chase\ source number 511 which is matched to cluster 04-6-013 from the ZKH08 catalog.

The ZKH08 catalog also contains a listing of background galaxies that were discarded from their cluster sample. We cross-correlated the X-ray source list with the ZKH08 galaxy list using the same criterion as above and found five matches (\chase\ source numbers 15=Zea20-6-037, 24=Zea21-6-004, 46=Zea26-3-007, 105=Zea30-6-023, and 433=Zea25-1-011).
Of those, \chase\ sources 46 and 105 show near coincidences with background galaxies in the HST images. Six non-GC sources in the SM07 catalog were matched to our X-ray sources (\chase\ sources 197=SMS107, 236=SMS133, 326=SMS213, 345=SMS230, 444=SMS356, and 455=SMS362), all of which were listed as being of an unknown nature. Source 345 is clearly a background galaxy, and number 444 and 455 both coincide with the nuclei of background galaxies.

\begin{figure}[ht]
\centering
\includegraphics[width=11cm,clip=t,angle=-90]{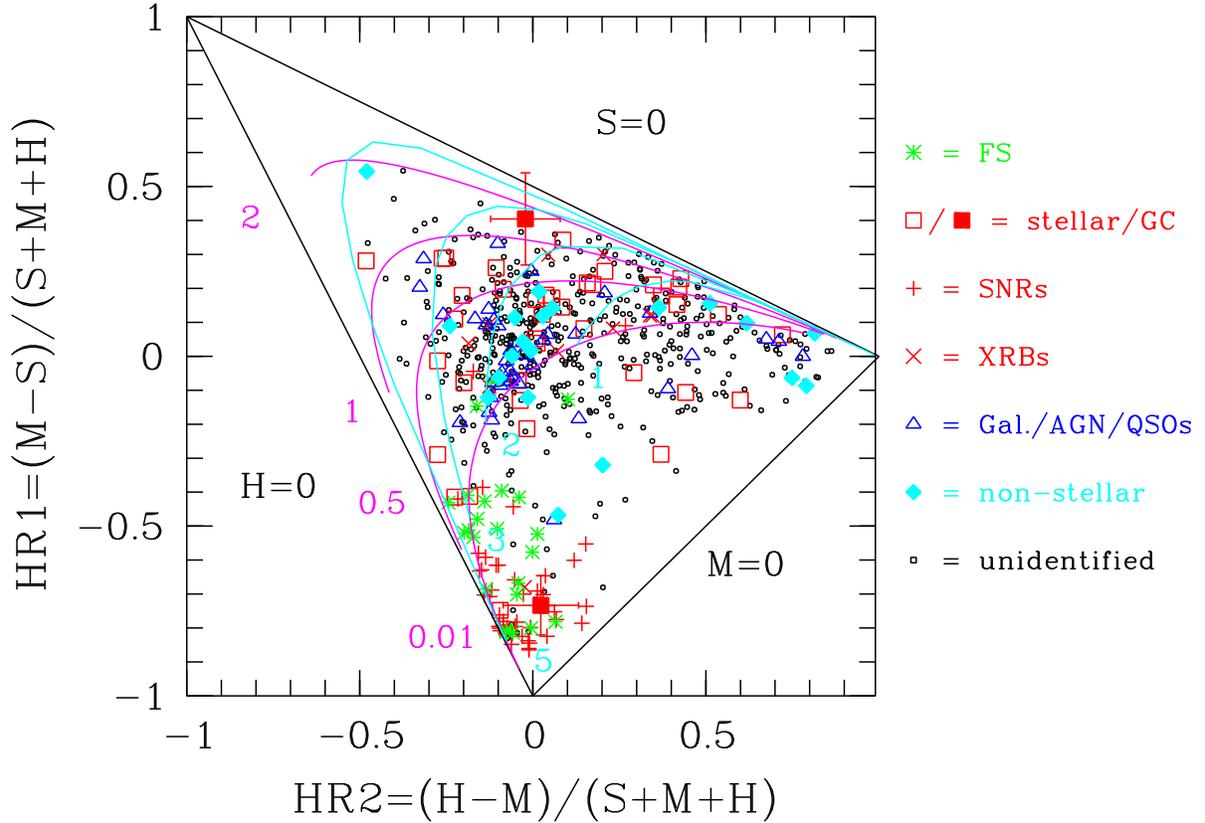}
\caption{\label{hr} Hardness ratio plot using a Bayesian approach \citep{park2006}. SNRs are from L10, whereas XRBs/candidates are taken from HP01, PMH04, MPH06, and W08. The magenta lines are trajectories for absorbed power-law models ($-1\le \Gamma \le6$) for which the \nh\ was fixed at 0.01, 0.5, 1.0, and $2.0\times10^{22}$\,cm$^{-2}$, while the cyan-colored trajectories assume a variable column density ($10^{20}$\,cm$^{-2}$$\le$\,\nh\,$\le$\,$10^{24}$\,cm$^{-2}$), but constant photon indices of 1, 2, 3, and 5. Representative error bars are shown for the two identified GCs.
}
\end{figure}
\subsection{Hardness ratios}\label{sec-hardness} 
We compute hardness ratios (HRs) using a definition similar to that described in \cite{prest09}, except that we extract source and background counts in the ``soft'' ($S$, 0.35\,--1.1keV), ``medium'' ($M$, 1.1\,--2.6keV), and ``hard'' ($H$, 2.6\,--8.0keV) energy bands.  These bands are identical to the ones used for the FLC and help to separate the soft sources (FSs and SNRs) from the hard sources (AGN).  Two different HRs are computed: $HR1 = (M-S)/(S+M+H)$, and $HR2 = (H-M)/(S+M+H)$.  We use a Bayesian approach which accounts for the fact that source and background counts are non-negative.  Although the {\em BEHR} code \citep{park2006} does not directly handle this form for the HRs, we adapt an approach described in a Chandra Source Catalog memo\footnote{http://cxc.harvard.edu/csc/memos/files/IEvans\_HardnessRatios.pdf} and evaluate the HRs based on the merged AE data for each source.

For each energy band, the inputs are source region counts, background region counts, a factor accounting for the ratio of the source and background extraction areas and efficiencies (AE's ``backscale'' parameter), and a factor converting from counts to flux (exposure time multiplied by the mean of the ARF over the extraction region). The BEHR code is run twice, once for $S$ and $M$, and again for $S$ and $H$; an option is set to save the ``draws'' for the posterior probability distributions of the two fluxes.  We set the ``burnin'' parameter to 50000 and the ``total draws'' parameter to 100000. This provides 50000 samples from the probability distributions for the fluxes. Combining the results from the two BEHR runs, we obtain 50000 samples from the probability distributions for the $S$, $M$, and $H$ fluxes. Using these distributions, we obtain 50000 values for HR1 and HR2. The HR value is obtained as the mean of the distribution, and the credible interval is evaluated based on the 68.2\% equal-tails estimates (i.e., 0.682/2 of the samples have values below the lower limit, and 0.682/2 of the samples have values above the upper limit).

In Fig.~\ref{hr} we plot the HR2 values vs. the HR1 values for all of the sources.  The limiting photon fluxes are indicated by solid black lines (the triangle): $S = 0$ (upper left to right center), $M = 0$ (right center to bottom center), and $H = 0$ (bottom center to upper left). Two distinctive regions with an enhanced source density are visible: one which contains the bulk of (mostly unidentified) sources around $(HR2,HR2) \sim (0.0, 0.1)$, and one at about $(HR2,HR1) \sim (-0.1,-0.8)$, following the $H=0$ line.  The first clump of sources near the center of the diagram appears to be a combination of XRBs, stellar and non-stellar sources, and background AGN. These objects appear to have \nh\ values between 0.01 and $0.5\times10^{22}$\,cm$^{-2}$ and photon indices between 1.0 and 2.5. The majority of the unidentified sources in our survey are in this clump, indicating that most of them are likely to be XRBs or AGN.

 The stretch of sources in the second clump with $HR1<-0.5$ along $H=0$ seems to be predominantly occupied by the L10 SNRs and foreground stars which tend to have significantly softer spectra.  An estimate of the \nh\ or the photon index for sources located around $(HR2,HR1)=(-0.1;-0.8)$ and $(HR2,HR1)=(0.5;0.2)$ is not possible as the model solutions in these regions become degenerate.

\subsection{Time variable sources in \chase}\label{subsec-variable}
The full \chase\ source catalog was systematically searched for sources which show significant long and short-term variability in the 0.35\,--\,8.0keV energy band. 
To check for variability, we applied a $5\sigma$ variability threshold, defined as $\eta=(flux_{max} - flux_{min})/\Delta flux$ across all observations (long-term variability) and a Kolmogorov-Smirnov (KS) test probing the probability $\xi$ of the source being constant within a single observation (short-term variability).  The KS test is part of AE and compares a uniform count rate model to the distribution of source event time stamps. The flux error $\Delta flux$ is calculated using the Gehrels approximation \citep{gehrels86} which is more appropriate for low count data.
To see how much the sources vary in flux, we calculated a variability factor defined as $flux_{max}/flux_{min}$, i.e. the ratio of the maximum flux to the minimum flux for a given source.

Conservatively, we consider any source with $\eta\geq5$ or $\xi\leq$\,5.7$\times$10$^{-7}$ (5-$\sigma$ equivalent) to be variable. We only used KS test results from observations when the source had $>$20 counts and was $<$8$'$ off-axis. In determining $\eta$, we only used observations where the source was $<$8$'$ off-axis, in order to avoid problems with background contamination where the PSF is very large. From our detailed spectral analysis (see Section~\ref{sec-specana}) we identified two additional sources at off-axis angles $>$8$'$ which also show time variability. These two sources are \#561 (X-9a) and \#612 (X-10).  

To test the robustness of our variability thresholds, we used the L10 SNR sample.  Since SNRs should not be time-variable, these tests provide a check against spurious variability detections.  As a result, the SNR with the highest long-term variability, L10-039, has a variability index of $\eta$=4.2 while the KS test for the SNR with the highest likelihood for short-term variability, L10-081, yields a $\xi$ of 0.025. These values are well outside of our high-confidence variability threshold.

The results of our variability analysis are shown in the column named 'Variability' of Table~\ref{crossref}, where high-confidence variables are shown in bold typeface. According to our conservative criteria, 38 sources are time-variable, 35 (7) are variable on short (long) time scales, 4 show both types of variability.  We note that $\eta$ and $\xi$ remain undefined if there are not enough observations in which a source has $>$20 counts (KS test minimum) and/or is $<$8$'$ off axis (KS and variability index maximum). The variability factor $flux_{max}/flux_{min}$ remains undetermined if the faintest observation had a flux $\le0$ or where there were fewer than 2 reliable flux measurements. 
If we are less conservative and set our variability criteria to values just outside of the range covered by the SNR sample ($\eta$\,$>$\,4.2 and $\xi$\,$<$\,0.025), 73 sources are time-variable, 49 (42) are variable on short (long) time scales, and 17 sources show both. 
We want to point out that the variability analysis is not optimized to find transient sources. Because the transients in \m\ are typically faint, can be far off-axis, and can sometimes only be detected in the unmerged observations \citep{ben08}, such transient sources may be too faint to pass our variability criteria. Running the transient identification routine from \citet{ben08} on the full catalog yielded 2 new transient candidates, 013345.24+304135.0 and 013420.91+303319.0.

\subsection{Supersoft X-ray sources}\label{sec-sss}
Although the front-illuminated ACIS-I chips are not the ideal choice to search for supersoft sources, we checked whether we detected any of the 12 SSSs reported by PMH04 and MPH06. It turned out that three sources are outside the \chase\ FOV, three are located at the border of the FOV and are not detected (faint {\it XMM} sources), while the remaining six sources are well within the \chase\ FOV but are not detected. Among these six objects one source is detected with \xmmnewton\ but is located in a region of diffuse emission and is most likely not a point source, while PMH04 source 247 (source 207 in MPH06) is a transient source named XRT-6 \citep[][W08 hereafter]{ben08} which is only detected in an ACIS-S observation (ObsID 786) which was intentionally excluded from our analysis. Table~\ref{crosss} provides a brief summary of the cross-identified SSSs. 

The seven sources from MPH06 that are not reported in PMH04 are sources that are only detected by MPH06 as faint sources in one observation of the {\it XMM} \m\ raster. PMH04, however, analyzed integrated images of the entire raster which are less sensitive for this type of source (due to the higher background) and therefore might have missed these sources. Besides lower sensitivity of ACIS-I for SSSs, the transient behavior of the sources might also be an additional reason why none of the {\it XMM} SSSs are detected by \chandra. 

To see if we can detect new SSSs besides the ones reported by PMH04 and MPH06, the entire \chase\ catalog was searched using criteria similar to the ones introduced by \citet{kong02}. We used the soft hardness ratio defined as $HR1=(m-s)/(s+m+h)$ based on $net\_cnts$ in the bands s=0.35\,--\,1.1keV, m\,=\,1.1\,--\,2.6keV, h=2.6\,--\,8keV and selected all sources with a hardness ratio $HR1$$<$$-0.5$. Among the 11 sources which passed the filter, eight sources are identified as L10 SNRs, two with stars, and one source is most likely a patch of diffuse X-ray emission in the giant \hii\ region NGC\,604. We detected no new SSSs in our sample.

\subsection{X-ray Spectral Analysis}\label{sec-specana}
Spectra were extracted by \ace\ using the {\tt CIAO} tool {\tt dmextract}. Response products were created by \ace\ using the {\tt CIAO} tools {\tt mkacisrmf} and {\tt mkarf}.  For sources observed in more than one observation, summed spectra were created and weighted response products were created by weighting by the exposure times for the individual observations and using the {\tt ftools} {\tt addrmf} and {\tt addarf}. The spectra were grouped using the grouping algorithm in \ace\ which allows the user to specify an energy range over which the grouping should be performed (0.35--8.0~keV in our case) and to specify a minimum signal-to-noise threshold to be achieved in each spectral bin in the net (background-subtracted) spectrum.  We specified a minimum signal-to-noise threshold of 2.0 for each spectral bin and a minimum of 8 spectral bins for a spectrum to be considered for spectral fitting. 254 of the 662 sources had a sufficient number of counts to satisfy both of these criteria.  The majority of these sources (163 of 254) had a minimum signal to noise ratio of 3.0 or larger in each spectral bin and 8 or more spectral bins, thereby approaching the regime in which Gaussian statistics are a good approximation. 
Given this grouping scheme, we used the reduced $\chi^2$ statistic as a measure of the goodness of fit. 

The grouped spectra for 254 of the 662 sources were automatically fit in \ace\ with an absorbed power-law model and, alternatively, with an absorbed $apec$ model \citep{smith01}. The Galactic absorption was modeled with the $tbabs$ model \citep{wilms00} by freezing the Galactic \nh\ at $6\times10^{20}$\,cm$^{-2}$ \citep{dickey90}. For the absorption component internal to \m, we assumed a $tbvarabs$ model with relative elemental abundances for elements heavier than He set to 0.5 (see \citet{rosolowsky2010} and references therein). This component was allowed to vary during the fit. In order to determine the best-fit model, models which predicted unreasonably high photon indices ($\Gamma >4.0$) and plasma temperatures ($kT>6.0$\,keV) were rejected. 
To distinguish between a power-law and a thermal plasma model, we checked which model provided the best fit to the spectrum (e.g., if there were indications for emission features, the thermal model was adopted). If both model fits appeared to be reasonable (a common occurrence  for sources with only a few hundred counts), the one with the lower reduced $\chi^2$ value was chosen. The best-fit model and parameters are listed in Table~\ref{bestfit}.
  
Because these simplistic single component models sometimes provide a poor fit, we also used multi-component models for sources which have a sufficiently large
number of counts to warrant fitting with more detailed spectral models.  There are 15 sources in the \chase\ catalog which have more than 2,000 net counts in the 0.35\,--\,8.0~keV band.  Most of these sources (13/15) are the well-known \einstein\ sources, including of course the nucleus of \m, which has the largest number of counts. Four of these sources have already been fit with more complicated spectral models and the results were published in earlier papers from the \chase\ project \citep[see][L10]{pietsch06,gaetz07}.   For the remainder, we examined the one component fits for the sources looking for structure in the residuals which might suggest a better, more complicated model.  We experimented with more complex models which added additional components such as an accretion disk model ({\tt   diskbb} in {\tt XSPEC}) and/or multiple thermal components.  For some of the sources we were able to identify a model which improved the fit significantly compared to the single component power-law or thermal models.  Below, we comment on the sources with the largest number of net counts, moving from highest number of counts to lowest number of counts. 

{\bf 013350.89+303936.6 (X-8, the Nucleus)}: The nucleus is the brightest X-ray source in \m\ and has been suggested to be a binary system containing an accreting stellar mass black hole ($\gtrsim 5$M$_{\odot}$) with a variable accretion disk \citep[e.g.,][]{dubus02,lapa03,wewa09}. Spectral analysis of the nucleus requires special care as the source is so bright that pileup is a significant concern.  The F1e1, F1e2, and ObsID 1730 data are not useful for spectral analysis because the nucleus is severely piled-up distorting the spectral shape. The other \chase\ pointings were arranged when possible to place the nucleus near the edge or just off of the CCD to eliminate the transfer streak from affecting the data in the region of interest. However, for two observations, F4e2 and F6e1, the nucleus was on the chip and far enough from the chip edge so that the source did not move on and off the CCD as the satellite dithered.  The source was 9\farcm0 and 9\farcm8 off-axis respectively in the F4e2 and F6e1 observations which reduced the effects of pileup. There are 163,938 counts in the combined spectrum.  

The one component fits are poor; the thermal model resulted in a reduced $\chi^2$ of 2.93 and the non-thermal model resulted in a reduced $\chi^2$ of 3.40. \citet{dubus02} fit the S3 spectrum of X-8 from ObsID~787, in which the source is positioned 7\farcm7 off-axis, with an absorbed disk blackbody model.  \citet{lapa03} analyzed the S3 data from ObsID~2023 in which X-8 was 12\arcmin\/ off-axis and found that the fit improved with the addition of a power-law component, a thermal component, and a broad Gaussian centered at 0.96~keV.  We fit the data with a model similar to that of \citet{lapa03}, specifically a thermal component ({\tt apec}), plus a Gaussian, plus a disk blackbody component ({\tt diskbb}), and a non-thermal component ({\tt pow}) model.  In our initial fits the {\tt apec} component was unconstrained and went to unreasonably low temperatures and high normalizations.  We therefore set the temperature of the {\tt apec} component to the value determined by \citet{lapa03} of kT=0.18\,keV since they analyzed data from the BI CCD (S3) which is more sensitive at energies below 1.0\,keV.  This fit resulted in a reduced $\chi^2$ of 1.02 for 237 degrees of freedom (DOF). The results are summarized in Table~\ref{11fitspect} for this fit and the other sources. The thermal component has a low normalization and contributed only to the lowest few channels. The vast majority of the flux is contained in the disk blackbody and power-law components.  Our fit resulted in an central energy for the Gaussian of $1.15^{+0.05}_{-0.06}$~keV which is somewhat higher than the value which \citet{lapa03} derived, $0.96^{+0.03}_{-0.10}$~keV. Future observations of X-8 with high-resolution spectrometers would be useful to characterize this feature.  
 
We examined the F4e2 and F6e1 spectra independently to search for variability between the epochs.  We fit each spectrum with the model described above and determined that the fit values of all of the parameters were consistent with each other at the 90\% confidence level.  The normalization of the disk blackbody component was 20\% higher in F4e2 than in F6e1, but this was only significant at the $1\sigma$ confidence level. These different normalizations for F4e2 and F6e1 imply values for the inner radius of the accretion disk of $R_{in}(cos\theta)^{1/2}=62.9$\,km in F4e2 and $R_{in}(cos\theta)^{1/2}=57.4$\,km in F6e1. The large range of allowed values for the parameters is partially due to the complexity of the multi-component model we have adopted in which some parameters can be strongly coupled to other parameters.  The best-fit value of the power-law index varied from 0.77 for F4e2 to 1.18 for F6e1, but again this difference was only significant at the $1\sigma$ confidence level. Although there is a clear variation in the total flux of X-8 between these two observations, it is not possible to ascribe the variability to just one component in our assumed spectral model. The power-law index for the combined observations is $\Gamma=1.20^{+0.29}_{-0.40}$. Our results are consistent with X-8 being an accreting black hole with a variable accretion disk. The combined spectrum is displayed in Figure~\ref{11fitspecf} together with those for the other sources.

{\bf 013334.13+303211.3 (X-7)}:  Detailed spectral fits and a timing analysis of X-7 have been carried out by \citet{pietsch06}.  The source is an eclipsing black hole X-ray binary \citep[see also][]{oro07} and was fit with a {\tt diskbb} model.

{\bf 013328.69+302723.6 (X-6)}:  There are a total of 47082 counts in the combined spectrum of X-6 from  F5e1, F5e2, F6e1, and ObsID 7208. The one component fits are poor; the thermal model resulted in a reduced $\chi^2$ of 1.41 and the non-thermal model resulted in a reduced $\chi^2$ of 1.58.  We fit this spectrum with a disk blackbody and a power-law model which resulted in a small improvement in the reduced $\chi^2$ to 1.35 with 242 DOF. This fit is still formally unacceptable but we were unable to find any other model combinations which fit the data better.  The source showed some evidence of variability in the \chase\ data sets, the long-term variability index $\eta$ is 3.9 which is below our threshold of 5.0 to be classified as confidently variable.  There was no evidence for short-term variability. The long-term variability may be part of the explanation for the relatively poor fit to the combined spectrum. We fit the F5e1, F5e2, and F6e1 spectra separately since they had between 13706 and 16108 counts, but we did not fit ObsID 7208 separately since it only had 1899 counts.  The disk blackbody plus power-law model results in better fits to the F5e1, F5e2, and F6e1 spectra with reduced $\chi^2$ values of 1.17, 1.11, and 1.07 respectively. All of the fit parameters are consistent with each other across the epochs at the $2\sigma$ confidence level.  X-6 is located in an uncrowded region of the galaxy with no obvious optical, IR, or UV counterpart.  We conclude that this source is most likely an XRB in \m\ since the disk blackbody model fits the spectrum best and there is a marginal indication of variability. Future studies to characterize the variability of this source in X-rays and to determine its optical counterpart are needed to confirm the classification as an XRB.    

{\bf 013324.40+304402.4 (X-5)}: There are a total of 33863 counts  in the combined spectrum of X-5 from  F1e1, F1e2, F3e1, F3e2, F4e1, F4e2, and ObsID 1730. The one component fits are poor; the thermal model resulted in a reduced $\chi^2$ of 1.53 and the non-thermal model resulted in a reduced $\chi^2$ of 1.58.  The fit with the disk blackbody and power-law model improved the reduced $\chi^2$ to 1.21 for 239 DOF. This source is clearly variable, as the count rate varied by $\sim33$\% between F3e1 and ObsID 1730. The long-term variability index $\eta$ is 8.8, but there is no indication of short-term variability.  We attempted to determine spectral differences between the various epochs but since there were only about $\sim5000$ counts in a spectrum for an epoch we were unsuccessful. Similar to X-6, this source is in an uncrowded region with no obvious optical, IR, or UV counterpart. We conclude that this source is most likely an XRB in \m\ since the disk blackbody model fits the spectrum best and there is convincing evidence for variability. 
    
{\bf 013451.85+302909.7 (X-10)}: There are a total of 26303 counts  in the combined spectrum of X-10 from  F6e1 and F6e2. The one component fits are acceptable; the thermal model resulted in a reduced $\chi^2$ of 1.01 and the non-thermal model resulted in a reduced $\chi^2$ of 1.04.  The fit with the disk blackbody and power-law model improved the reduced $\chi^2$ slightly to 0.99 for 236 DOF. There is significant evidence for both long-term and short-term variability ($\eta=12.5$ and $\xi=6.62\times10^{-10}$). X-10 is located in an uncrowded region on the eastern side of the galaxy near the border of the D$_{25}$ isophote.  There is no optical counterpart, but there is some faint emission in the 2MASS J band which might be a counterpart. We conclude that this source is an XRB in \m\ given its nonthermal X-ray spectrum and its short-term variability. 

{\bf 013315.16+305318.2 (X-4)}: There are a total of 23117  counts  in the combined spectrum of X-4 from  F3e1 and F3e2. The one component fits are poor; the thermal model resulted in a reduced $\chi^2$ of 2.22 and the non-thermal model resulted in a reduced $\chi^2$ of 2.03.  The fit with the disk blackbody and power-law model improved the reduced $\chi^2$ significantly to 1.13 for 239 DOF. There is no evidence for variability between these two epochs ($\eta$ is 0.6 and $\xi$ is $1.48\times10^{-2}$). We conclude that this source is most likely an XRB in \m, since the quality of the fit improved dramatically for the disk blackbody plus power-law model.   Future studies should try to detect and characterize any variability in X-rays and to determine the optical counterpart to confirm the classification as an XRB.
    
{\bf 013311.75+303841.5 (X-3)}: X-3 is the brightest SNR in \m\ (G98-21, see \citet{gordon98}) and a detailed analysis of the spectrum was presented in \citet{gaetz07}. 

\begin{figure}[!pt]
\centering
\includegraphics[width=\textwidth,angle=0,clip]{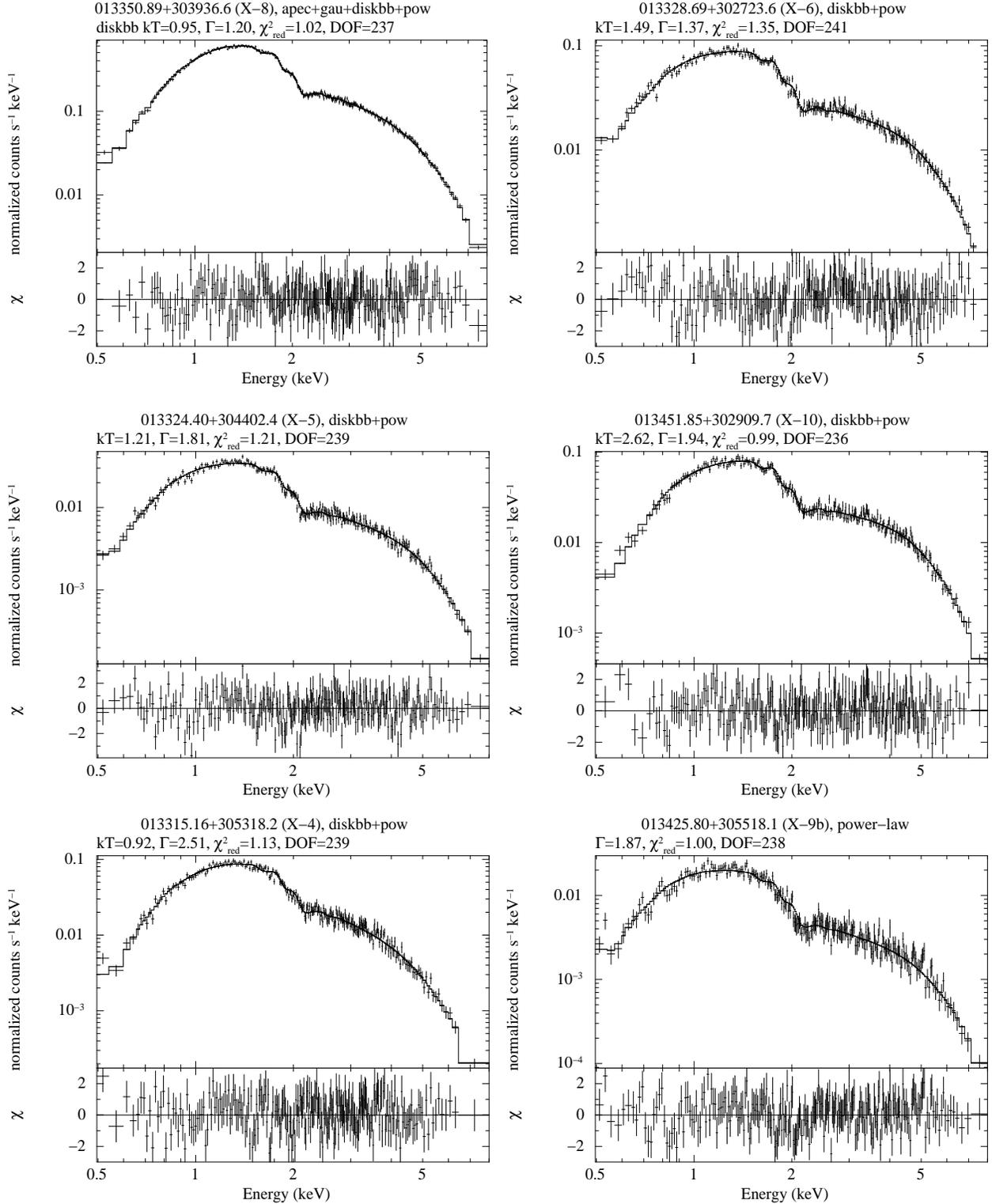}
\caption{\label{11fitspecf}
Spectra and refined spectral fits for the brightest \chase\ sources. The best-fit model is either a disk blackbody plus non-thermal model ({\tt diskbb+pow}) or a simple power-law model ({\tt pow}), except for the nucleus, which requires a third component ({\tt apec}) in addition to the {\tt diskbb} and {\tt pow} components. The column density is modeled by a two component absorption model, consisting of the Galactic column density and the one internal to \m.}
\end{figure}

\begin{figure}[!pt]
\centering
\addtocounter{figure}{-1}
\includegraphics[width=\textwidth,angle=0,clip]{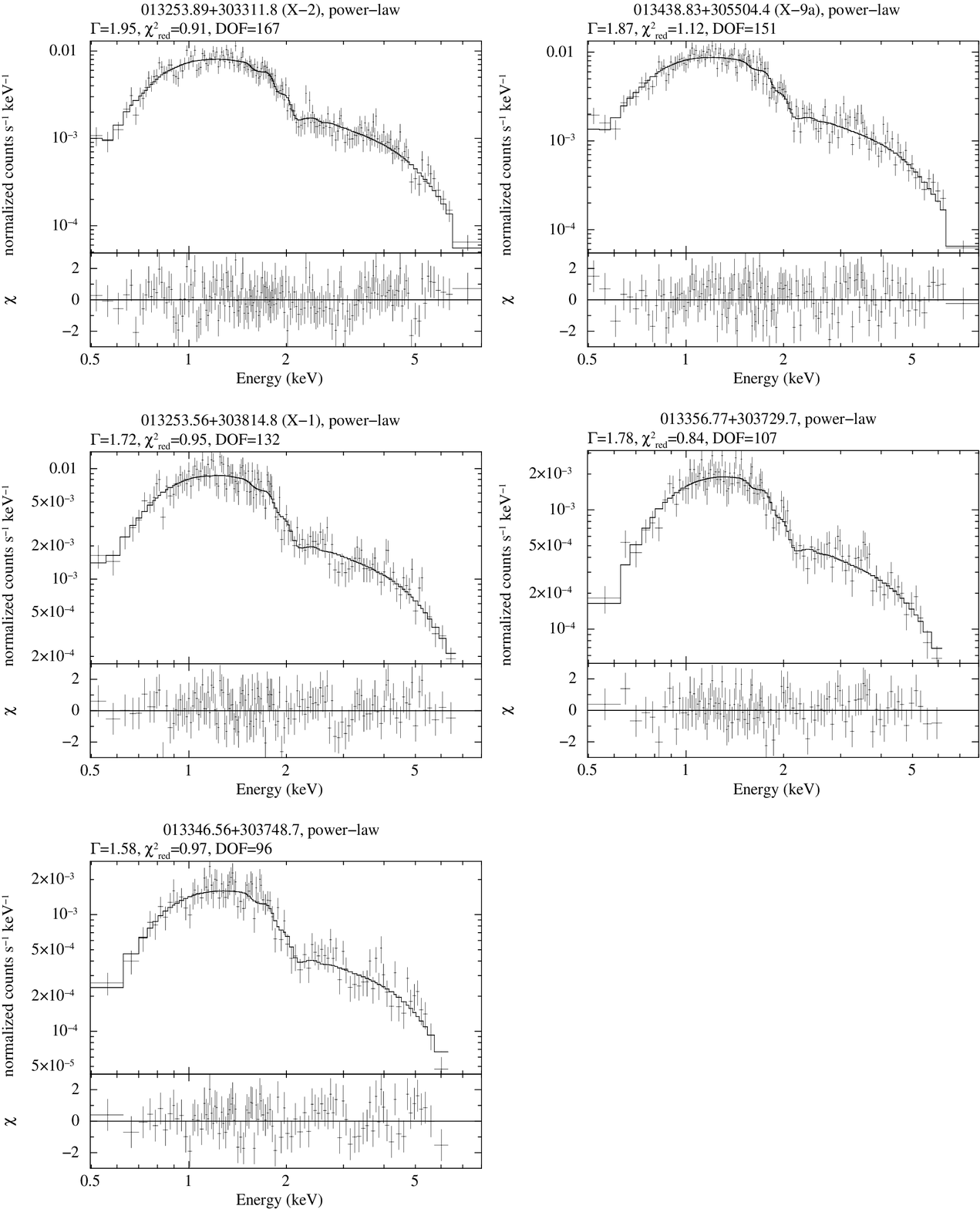}
\caption{Continued.}
\end{figure}

{\bf 013425.80+305518.1 (X-9b)}: There are a total of 9064 counts in the combined spectrum of X-9b from  F2e2, F2e1, and ObsID 2023. The one component thermal model resulted in a reduced $\chi^2$ of 1.15 and is unacceptable, but the one component nonthermal model resulted in a reduced $\chi^2$ of 1.00 and is acceptable. The two component disk blackbody and power-law model also resulted in a reduced $\chi^2$ of 1.00 (for 236 DOF) and therefore does not significantly improve the fit. Based on these results, we conclude that the power-law model is the best fit to the data.  The source is highly variable with the flux changing by a factor of three between F2e2 and ObsID 2023. The long-term variability index $\eta$ is 38.7. Unfortunately, there are not enough counts in a given epoch to search for spectral variability between the epochs (1943 counts in F2e2 and 4578 counts in ObsID 2023). The source has a clear counterpart in the optical and the IR. It is bright in all of the \spitzer\ and 2MASS bands.  We conclude that this source is most likely an AGN or an XRB since the non-thermal model fits the spectrum best with a spectral index of 1.87 typical of AGN and XRBs, there is evidence for long-term variability, and the source has a bright optical and IR counterpart.  Followup optical spectroscopy would be useful in distinguishing between an AGN and XRB.

{\bf 013253.89+303311.8 (X-2)}: There are a total of 4891 counts in the combined spectrum of X-2 from  F4e1, F4e2, F5e1, and F5e2. The one component fits are acceptable; the thermal model resulted in a reduced $\chi^2$ of 1.05 and the non-thermal model resulted in a reduced $\chi^2$ of 0.91.  The fit with the disk blackbody and power-law model only improved the reduced $\chi^2$ to 0.90 for 165 DOF.  Based on these results, we conclude that the power-law model is the best fit to the data.  The source is variable with the flux changing by $\sim30$\% between F4e2 and F5e1 ($\eta$ is 7.2). Unfortunately, there are not enough counts in a given epoch to search for spectral variability between the epochs (1430 counts in F4e2 and 1167 counts in F5e1). The source has a bright optical counterpart for which we acquired an MMT spectrum.  The optical spectrum is consistent with an AGN at a redshift of 0.49.  The optical source was also identified as variable in the Canada-France-Hawaii Telescope (CFHT) survey of \m\/ \citep{hart06} as source CFHT~250850. We conclude that this source is certainly an AGN given its non-thermal spectrum with a spectral index of 1.95, long-term variability in X-rays, and its optical spectrum and redshift of $z=0.49$.

{\bf 013438.83+305504.4 (X-9a)}: There are a total of 4044 counts in the combined spectrum of X-9a from  ObsID 2023, F2e2, and F2e1. The one component fits are reasonable, but not formally acceptable. The thermal model resulted in a reduced $\chi^2$ of 1.15 and the non-thermal model resulted in a reduced $\chi^2$ of 1.12.  The fit with the disk blackbody and power-law model resulted in no improvement in the reduced $\chi^2$ (1.12 for 149 DOF).  Based on these results, the power-law model is the best fit to the data. The source is clearly variable as the long-term variability index $\eta$ is 11.9, but there is no evidence for short-term variability. The source has a bright counterpart in the optical, IR, and UV.  Our MMT spectrum of the source is consistent with an AGN at a redshift of $z=0.89$.  We conclude that this source is certainly an AGN given its non-thermal X-ray spectrum with a spectral index of 1.87, X-ray long-term variability, and redshifted spectrum.

{\bf 013331.25+303333.4 (X-14)}: X-14 is the second brightest SNR in \m\ (G98-31, see \citet{gordon98}) and a detailed analysis of the spectrum was presented by L10.

{\bf 013253.56+303814.8 (X-1)}: There are a total of 2808 counts in the combined spectrum of X-1 from  F4e1 and F4e2. The one component fits are acceptable; the thermal model resulted in a reduced $\chi^2$ of 0.97 and the non-thermal model resulted in a reduced $\chi^2$ of 0.95.  The fit with the disk blackbody and power-law model resulted in no improvement in the reduced $\chi^2$ (0.96 for 130 DOF).  Based on these results, the power-law model is the best fit to the data.  The source showed only marginal evidence for long-term variability ($\eta$ is 4.3).  The source has a bright counterpart in the optical and IR.  Our MMT spectrum of the source is consistent with an AGN at a redshift of $z=0.37$.  The source is also identified with the optical variable CFHT~250829. We conclude that this source is certainly an AGN given its non-thermal X-ray spectrum with a spectral index of 1.74, redshift, and variability in the optical.

{\bf 013356.77+303729.7}: There are a total of 2434 counts in the combined spectrum of 013356.77+303729.7 from ObsID 1730, ObsID 7208, F1e1, F1e2, F5e1, F5e2, F6e1, F6e2, and F7e2.  This source was not detected by \einstein\/.  The combined exposure in the \chase\/ survey is over 700~ks, so it is perhaps not surprising that \chase\/ detected this source while \einstein\/ did not. The one component fits are acceptable; the thermal model resulted in a reduced $\chi^2$ of 0.83 and the non-thermal model resulted in a reduced $\chi^2$ of 0.84.  The fit with the disk blackbody and power-law model resulted in no improvement in the reduced $\chi^2$ (0.85 for 105 DOF).  Based on these results, we have a slight preference for the thermal model. However, the source is clearly variable as the flux varied from ObsID 1730 to F1e1 by a factor of 6 ($\eta$ is 5.7). Based on the variability, we have a slight preference for the non-thermal model. 
The source is located in a crowded region, so the optical counterpart is not secure. 
There is a variable source, CFHT~235490, inside the error circle for the X-ray source.  We conclude that this source is most likely an AGN or an XRB since the non-thermal model fits the spectrum as well as the thermal model, the best-fit spectral index is 1.82, and there is convincing evidence for long-term variability.

{\bf 013354.91+303310.9 (X-13)}: X-13 is the third brightest SNR in \m\ (G98-55, see \citet{gordon98}). A detailed analysis of the spectrum was presented by L10.

{\bf 013346.56+303748.7}: There are a total of 2195 counts in the combined spectrum of 013346.56+303748.7 from ObsID 1730, ObsID 7208, F1e1, F1e2, F4e2, F5e1, F5e2, F6e1, and F6e2.  This source was not detected by \einstein\/.  The combined exposure in the \chase\/ survey is over 700~ks, so it is perhaps not surprising that \chase\/ detected this source while \einstein\/ did not. The one component fits are acceptable; the thermal model resulted in a reduced $\chi^2$ of 0.99 and the non-thermal model resulted in a reduced $\chi^2$ of 0.97.  The fit with the disk blackbody and power-law model resulted in no improvement in the reduced $\chi^2$ (0.99 for 94 DOF).  The source is clearly variable as the flux varied from F5e1 to F1e2 by a factor of 3 ($\eta$ is 11.8). Based on the variability, we have a slight preference for the non-thermal model. The source is close to the nucleus of the galaxy and is located in a crowded region.  A secure optical counterpart cannot be identified.  We conclude that this source is most likely an AGN or an XRB since the non-thermal model fits the spectrum as well as the thermal model, the best-fit value for the spectral index is 1.61, and there is convincing evidence for long-term variability.

\subsection{The X-ray Luminosity Function of \m}\label{sec-lumfun} 

\subsubsection{Conversion of Photon Flux to Energy Flux}\label{subsec:convert.photflux.to.eflux}
Both the radial profile determination and the $\log N$--$\log S$ relations discussed below require estimates for the energy flux, $S$, of a source. \ace\ provides a photon flux estimate (the {\it flux2\/} column) based on the source {\it net\_cts}, exposure, and the mean ARF in the given energy band (assuming a spectrum with a flat photon index).  To obtain the energy flux, some information about the intrinsic spectrum of the object is required. There are multiple ways of measuring the flux in a given energy band that could be used.  One could use fluxes calculated from the raw count rate using a single ad hoc spectral model (which can be done for all of our sources), or one could use fluxes which are calculated from the raw count rate using a simple spectral model derived from the hardness ratios (which can be done for all but the dimmest sources).  Alternatively, one could use fluxes derived from spectral fits (which can be done only for the brightest third of our sources). Applying different count rate to flux conversions, particularly at different flux levels, can introduce biases and spurious features in the $\log N$--$\log S$ distribution. As a result, we decided to convert photon fluxes to energy fluxes using an absorbed powerlaw spectral model with $\Gamma=1.9$ and $\nh=6\times10^{20}$\,cm$^{-2}$, which is considered to be representative for all sources (standard model hereafter). 
This kind of model appears to be appropriate for XRBs in \m\ and background AGN which dominate at the lower flux levels.  The model would be inappropriate for really hard/soft sources (where we would overestimate/underestimate the flux).  The flux estimate could be off by as much as a factor of two if the spectrum is significantly different than the assumed one.  However, the standard model seems to be a good choice for the majority of sources so that the net effect is not that large and does not bias the $\log N$--$\log S$ relation systematically to higher or lower fluxes. 

\subsubsection{Sensitivity Maps}\label{subsec:sensitivity.maps}
\begin{figure}[t]
\centering
\includegraphics[width=16.5cm,height=8.9cm,angle=0,clip]{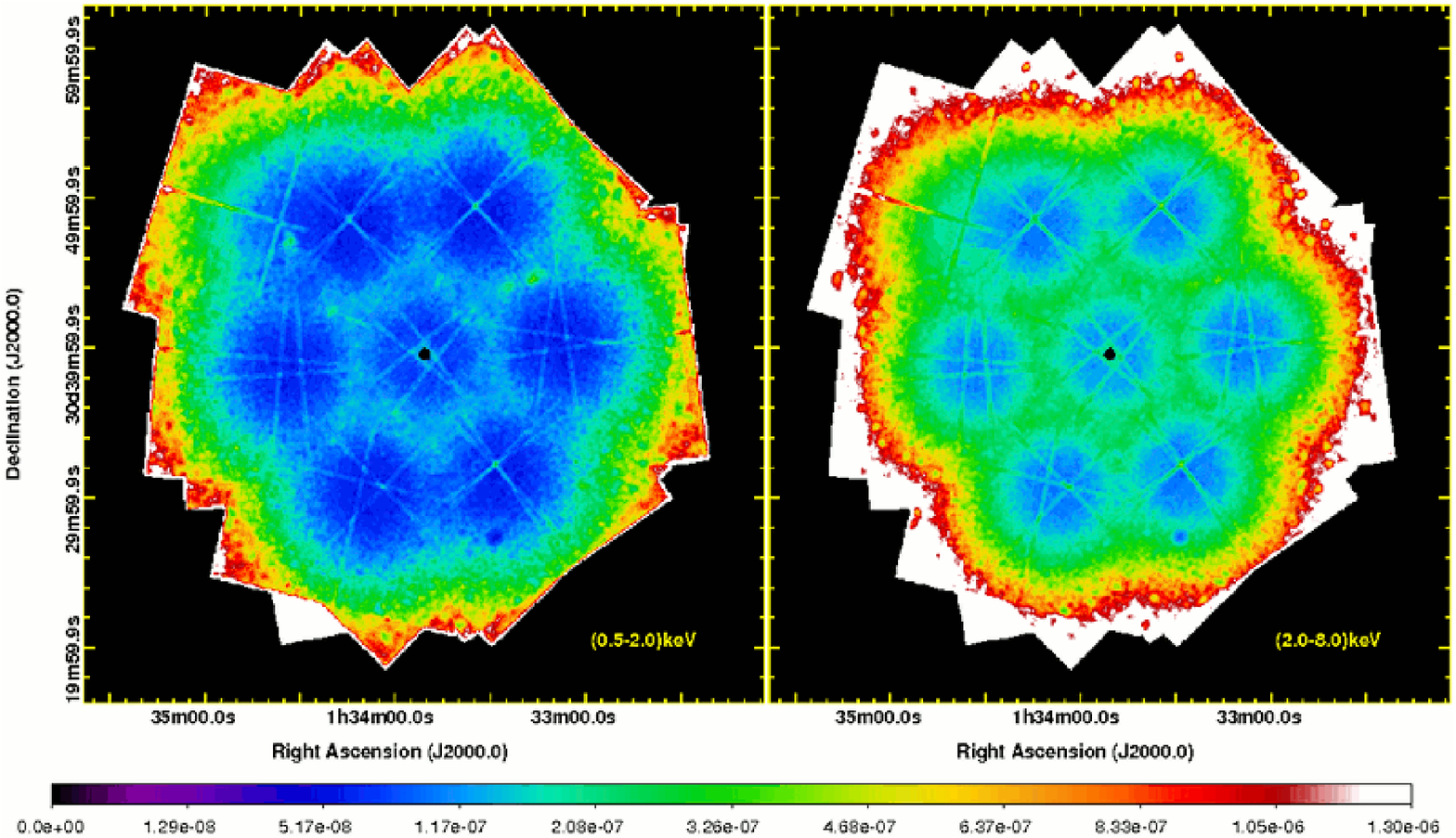}
\caption{\label{f6} Sensitivity map (in units of photons s$^{-1}$ cm$^{-2}$, square root scale) created in the 0.5\,--\,2.0\,keV (left) and 2.0\,--\,8.0\,keV energy band (right), assuming that each source can be represented by a power-law of $\Gamma=1.9$ and an average column density of $N_H=6\times 10^{20}$\,cm$^{-2}$. Since we excluded the region around the nucleus from our standard source detection and characterization, we set the exposure within a radius of 24\arcsec\ from the nucleus to zero.}
\end{figure}

For the following analyses, we require sensitivity maps which provide, for each point in the survey area, the energy flux level at which a source would be detectable.  Sensitivity maps consistent with the \ace\ selection criteria are, in the absence of a high density of sources, relatively straight-forward to generate.  For each point in the source image we calculated the size of the source extraction region and the corresponding local background rate.  From these quantities one can calculate the number of counts required within the exposure for which the $pns$ falls below $4\times10^{-6}$. Based on the standard model, the count rate is then converted to the corresponding energy flux and luminosity. Of course, multiple exposures and field centers are a complication. We combined all data for each field and calculated sensitivity maps for the individual fields. We then combined the sensitivity maps for all of the fields. If at a point in the region of overlap there are several different measures of the sensitivity, $[S_0,S_1,...,S_n]$, the resulting sensitivity was taken to be Min(Min($[S_0,S_1,...,S_n]$), ($\sum_N{S_i^2})^{1/2}$), i.e., either the best individual field sensitivity at that point, or the sensitivity of the summed data, whichever was smaller. 
Since we excluded the region around the nucleus from our standard source detection and characterization, we set the exposure within a radius of 24\arcsec\ from the nucleus to zero. The sensitivity limit of $10^{-7}$\,photons cm$^{-2}$ s$^{-1}$ corresponds to completeness limits of 34\%, 51\%, and 66\% for semi-major axes of 30\arcmin, 20\arcmin, and 10\arcmin, respectively. In other words, 34\% of the sky area of the sensitivity map within 30\arcmin\ has a value $\ge$$10^{-7}$\,photons cm$^{-2}$ s$^{-1}$ and likewise for the other semi-major axes mentioned
The resulting sensitivity maps for the standard model in the soft and hard energy band are shown in Figure~\ref{f6}. 

\subsubsection{Radial source distribution}\label{subsec:radial.source.dist}
Direct multi-wavelength detection of counterparts to the X-ray sources allows the classification of X-ray sources and provides a means to discriminate between sources in \m\ and those in either the foreground or background (see \S\ref{sec-comp}).  Such a detection scheme is limited in that some objects in \m\ or faint background sources, are unlikely to be detected and classified in such multi-wavelength studies. As a consequence, the nature of a significant number of sources will remain ambiguous. Statistical tests on the spatial distribution of the sources, however, can constrain directly the number of sources in \m, as well as providing useful information for the construction of the $\log N$--$\log S$ distribution and the LF of \m.

\begin{figure}[pht]
\centering
\includegraphics[width=\textwidth,angle=0,clip]{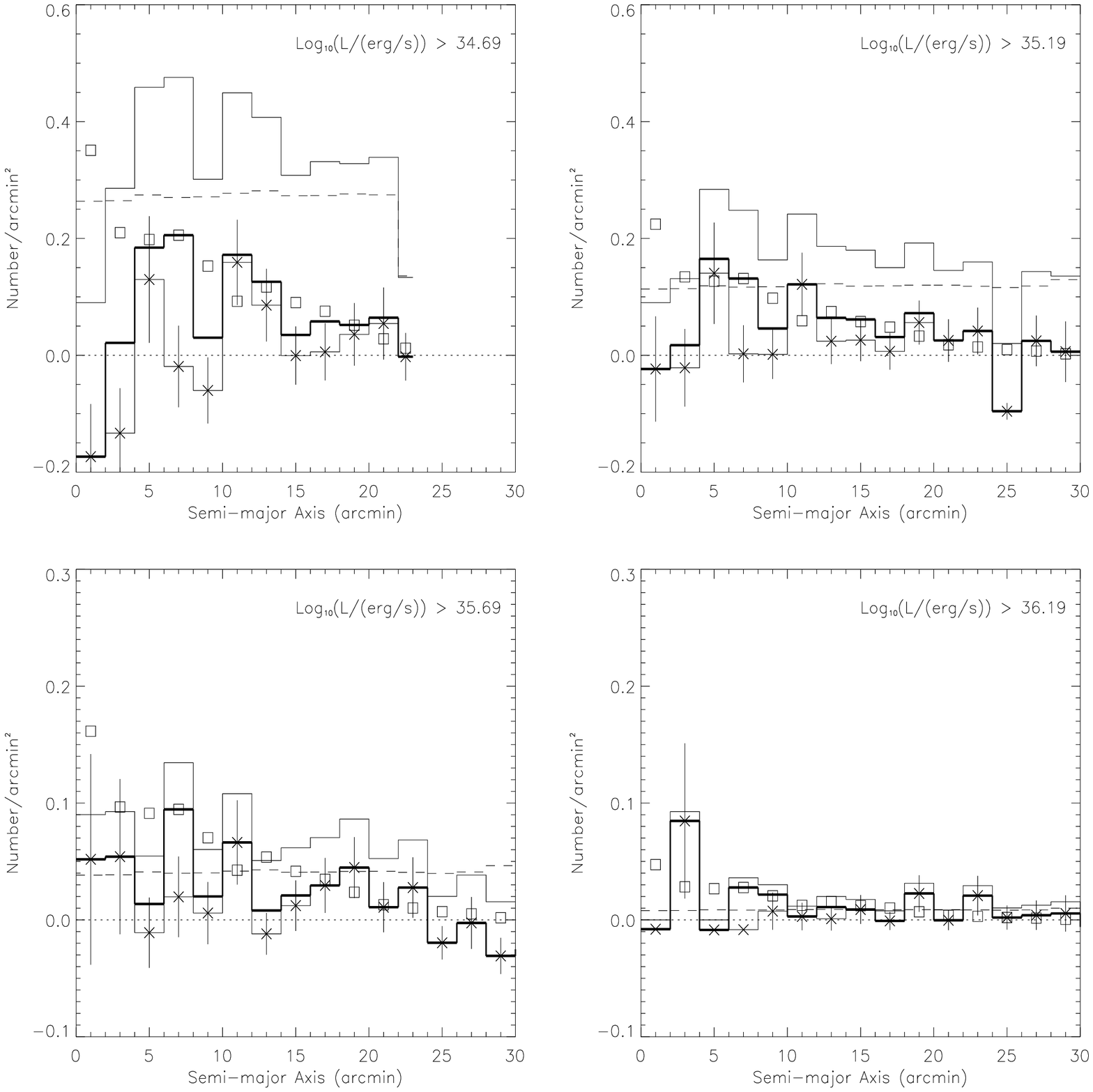}
\caption{\label{radprof} The galactocentric source density profiles in the 0.5\,--\,2.0\,keV energy band. In each panel, the {\it top histogram} represents the total number of sources greater than the indicated luminosity after the removal of foreground stars. The {\it dashed histogram} is the expected number of background AGN, the {\it thick histogram} addresses the number of sources after the removal of the expected number of background AGN, while the {\it bottom histogram (with error bars)} represents the number of sources after the removal of the expected number of background AGN and the removal of the known SNRs. {\it Open boxes} follow a scaled profile of the \galex\ FUV surface brightness.}
\end{figure}

The galactocentric profile was created by summing the number of sources within ellipses concentric with the $D_{25}$ contour within $\Delta\log(S)=0.1$ intervals, forming a two dimensional histogram.  Each bin of this histogram was divided by the area within the annulus over which sources within the flux interval could have been detected.  Histogram bins for which that area was less than one tenth of the total area of the annulus were eliminated from further use.  The resulting galactocentric profiles are shown in Figure~\ref{radprof} for four different luminosity levels. The only sources removed before these histograms were constructed were the foreground stars. They are few in number and it is unlikely that a significant number of them was missed.

At each flux level the raw number of sources generally decreases with galactocentric radius. We used the cumulative $\log N$--$\log S$ distribution of \citet{cap09} and a map of the total hydrogen nucleon column density (N(HI)+2XN(CO), where X\,=\,$2\times10^{20}$\,cm$^{-2}$) to calculate the expected number of AGN at each location after their flux has been reduced by foreground galactic absorption and absorption by the disk of \m. The expected contribution is shown by the dashed line in Figure~\ref{radprof}. In each case the expected contribution from background AGN is nearly flat at most galactocentric radii, with an increase at the largest radii at the edge of the \ion{H}{1} disk.

Subtracting the expected number of AGN from the total profile leaves the same downward trend with radius (the thick histogram in Figure~\ref{radprof}). The identification of \m\ SNRs has been well studied (e.g. by L10) though the completeness of their SNR catalog is not clear. As most of the SNRs included in our source catalog are point-like, the completeness for the set of SNRs in our catalog will be similar to that of the other point sources. After the removal of the SNRs, the profile becomes significantly flatter, though there is still a positive signal. Although the majority of sources are in \m, there could still be some AGN in those bins. 


Figure~\ref{radprof} reveals also some significant features at $\sim$7\arcmin\ and $\sim$12\arcmin. While the excess of sources at the smaller radius could correspond to prominent spiral arms tangent to the ellipses concentric with the $D_{25}$ ellipse, the feature at the larger radius seems to be due to aggregations of faint sources.

\subsubsection{The $\log N$--$\log S$ relation and the \m\ X-ray Luminosity Function}\label{subsec-lumfun} 
For a survey covering the total geometric area $A$, the cumulative number of sources $N(>S)$ can be evaluated by summing over all sources with fluxes exceeding $S$, weighted by the survey area, $A(S)$, over which a source with flux $S$ could have been detected:
\begin{equation}
N(>S) = \sum_{i, S_\mathit{i} > S} {1 \over A(S_\mathit{i})} .
\end{equation}

Here, $S$ is the energy flux in units of $\mathrm{erg}\,\mathrm{cm}^{-2}\,\mathrm{s}^{-1}$, and $A(S)$ survey area over which a source with flux $S$ could have been detected. $N(>S)$ and the corresponding differential form ($N(S)$) thus have units of sources per solid angle, or in our case, $\mathrm{sources}\;\mathrm{deg}^{-2}$.  Note that the $A(S)$ factor would account for survey completeness if the sources were distributed uniformly. Although the LF could vary, the number of sources is not sufficient to show a significant difference at low fluxes between the inner and outer halves of the area covered by the survey. As a practical matter, we terminate the low flux end of the distribution when $A(S)/A$ falls below 10\% of the total geometric area of the survey. The sensitivity maps calculated above (\S\ref{subsec:sensitivity.maps}; see Fig.~\ref{f6}) allow the area function, $A(S)$ to be evaluated, by summing the sensitivity map over the regions where a source with flux $S$ would satisfy our {\it pns\/} detection criterion. If $N(>S)$ is given, the differential version can be evaluated as: 
\begin{equation}
N(S)\,\Delta S = {N(>[S+\Delta S]) - N(>S) \over \Delta S}
\end{equation}
By itself, the cumulative or differential $\log N$--$\log S$ relation is of limited utility for the heterogeneous collections of \m\ objects mixed with foreground stars and background AGN.  Our aim is to dissect the $\log N$--$\log S$ distribution to extract the LF for the \m\ point source population.  Two components, the foreground stars and the SNRs in \m, are well in hand, and their contribution can be evaluated directly.  However, because \m\ subtends a large solid angle on the sky, the contamination by background AGN is a serious issue.  We start with the differential $\log N$--$\log S$ distribution, because the variation of the components is easier to discern.

\begin{figure}[hpt]
\centering
\includegraphics[width=\textwidth,angle=0,clip]{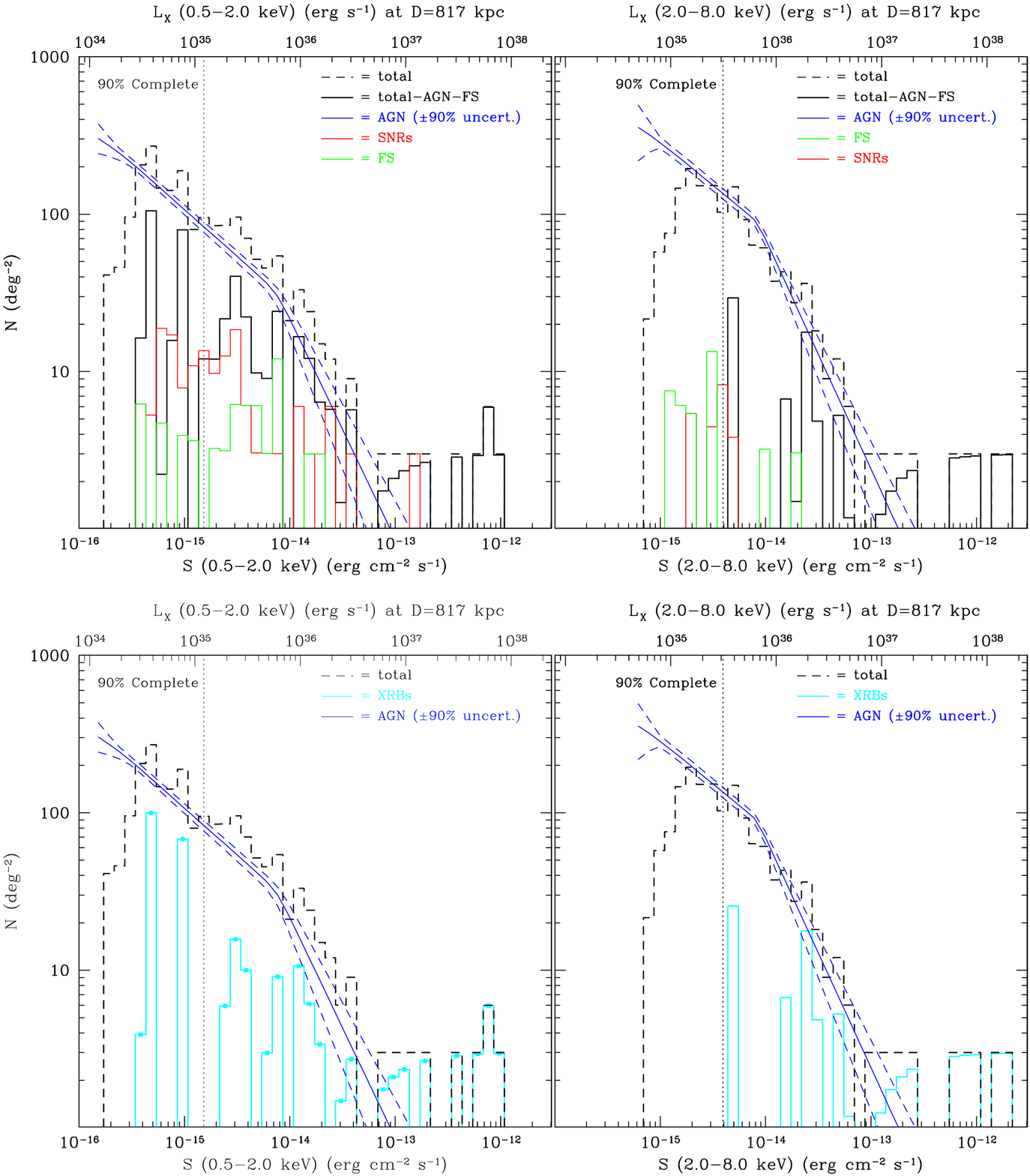}
\caption{\label{xlfdif} Differential X-ray $\log N$--$\log S$ for \m\ in the soft (0.5\,--\,2.0keV) and hard (2.0\,--\,8.0keV) energy band. The uncertainty of the AGN model from \citet{cap09} was evaluated for a 90\% Poisson error. The corresponding X-ray luminosities (assuming the objects are in \m) are shown on top of each panel.
}
\end{figure}

In Fig.~\ref{xlfdif}, we show the differential $\log N$--$\log S$ relation for the soft (0.5--2\,keV) and hard (2--8\,keV) bands in the left and right panels, respectively. We start by subtracting foreground stars and SNR contributions from the total source distribution, resulting in a combination of \m\ point sources and background AGN.

\begin{figure}[ht]
\centering
\includegraphics[width=\textwidth,angle=0,clip]{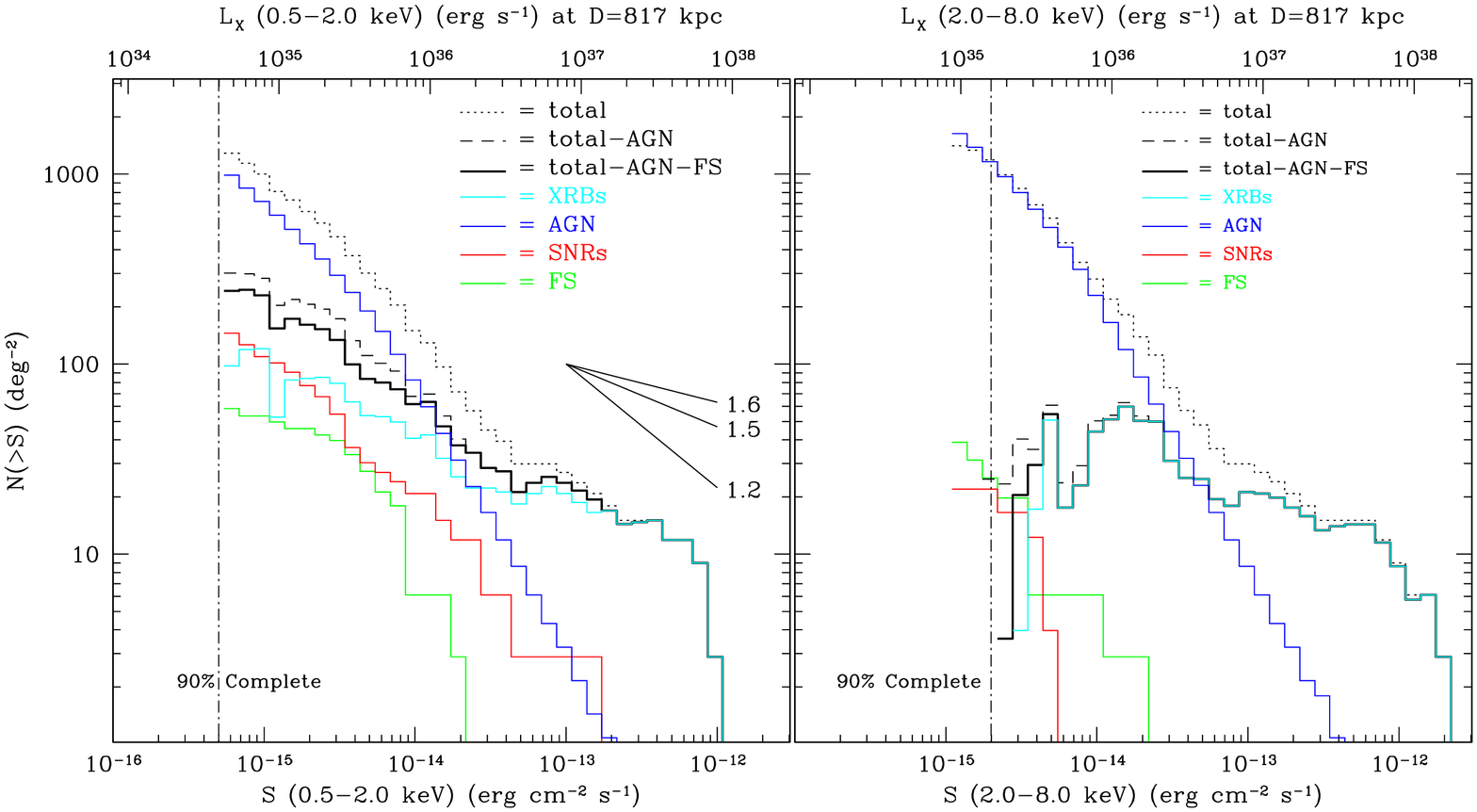}
\caption{\label{f7}
Cumulative X-ray $\log N$--$\log S$ for \m\ in the soft (0.5\,--\,2.0keV) and hard (2.0\,--\,8.0keV) energy band. For the AGN $\log N$--$\log S$, the model from \citet{cap09} was assumed. The solid black histogram corresponds to the sources we think are in \m, while the solid cyan line reflects the XRBs (after SNRs have been subtracted). The three black lines represent different cumulative slopes for the HMXBs.}
\end{figure}

Subtracting the AGN contribution requires some care, since the AGN dominate strongly at lower flux values. For each $1\arcsec$\,$\times$\,1$\arcsec$ pixel of our image, we calculated the differential $\log N$--$\log S$ relation of the background AGN.  We accounted for \m\ absorption by evaluating $e^{-\sigma N(H)}[S,S+\delta S)$, where N(H) is the total hydrogen (nucleon) column density (N(HI)+2XN(CO)) for that pixel.  For the AGN $\log N$--$\log S$, we adopted the distribution provided by \citet{cap09}, which nicely confirms the earlier work from \citet{harri03}. 

In Fig.~\ref{f7} we present the cumulative X-ray LF for all sources detected in the FOV corrected for absorption and contributions due to background AGN. We used vertical black lines to mark the flux for which $\int N(S)dS/\int N(S)(A/A(S))dS \sim 0.9$, i.e. the level at which the survey is $\sim$90\% complete. The cumulative $\log N$--$\log S$ for the AGN (blue line in Fig.~\ref{f7}) was obtained by summing the AGN differential $\log N$--$\log S$ in the same manner as the measured differential $\log N$--$\log S$. The cumulative $\log N$--$\log S$ for the AGN was then subtracted from the uncorrected $\log N$--$\log S$ to give the background-corrected cumulative $\log N$--$\log S$ for \m\ (black dashed line in Fig.~\ref{f7}). The cyan curve is what remains left after contributions from SNRs and FSs are removed, i.e., the distribution for sources in \m .  By converting the fluxes to luminosities (based on the assumed \m\ distance), the corresponding \m\ X-ray LF is obtained.

There are four sources of uncertainty in deriving the X-ray LF. The first uncertainty is due to Poisson statistics for X-ray source numbers which become increasingly important at higher luminosities and dominate over the other sources of uncertainty at most luminosities.  The second uncertainty is due to intrinsic source variability. \citet{zezas07} have shown that in the case of the Antennae, where there is a strongly time-varying population, the LF from a single epoch is statistically consistent with the mean of multi-epoch measurements. Our LF is formed from at least two epochs of observation and should therefore be a reasonable representation of the mean LF.  The third uncertainty is due to the sensitivity calculation. Although this uncertainty produces only small changes in $A(S)$, it produces a large change in the expected number of background AGN, because the number of AGN increases rapidly at low fluxes. This effect is insignificant at higher fluxes, but becomes significant below the flux at which AGN start to dominate ($S\lesssim 5\times10^{-15}$\,erg cm$^{-2}$ s$^{-1}$ in the 2.0-8.0 keV band).

The fourth uncertainty is due to the uncertainty in the AGN $\log N$--$\log S$ relation. The \citet{cap09} $\log N$--$\log S$ distribution was derived from a deep ($2 \mathrm{deg}^2$) field and is consistent with the deeper Chandra Deep Field-South (CDF-S). Given the depth and FOV of the CDF-S compared to \chase, the uncertainty in the shape of the AGN $\log N$--$\log S$ distribution should be a smaller effect than the Poisson uncertainty in the number of expected sources (i.e., the normalization) in the \m\ FOV. 
To determine the magnitude of the normalization uncertainty we used the \citet{cap09} parametrization of the AGN $\log N$--$\log S$ relation to simulate the expected number of AGN within the FOV as a function of flux.  At each flux level we determined the limits containing 90\% of $10^4$ Monte Carlo simulations.  We then propagated these limits through our model of the absorption due to \m\ to determine the uncertainty in the AGN subtraction due to the uncertainty of the normalization.  To determine the magnitude of the effect of the uncertainty in the shape of the AGN $\log N$--$\log S$, we fit the data from the \citet{cap09} cumulative $\log N$--$\log S$. We then found the 90\% uncertainty interval for each parameter individually, allowing all the other parameters to vary. The envelope formed by the $\log N$--$\log S$  derived from the individual parameter limits is smaller (by roughly a factor of two at most fluxes) than the envelope due to the normalization uncertainty. We show the effect of the normalization error in Figure~\ref{xlfdif} rather than the uncertainty due to the shape of the $\log N$--$\log S$. This relative uncertainty becomes larger at higher fluxes where the total number of AGN becomes insignificant.

From the cumulative $\log N$--$\log S$ we see that the source density of X-ray sources in \m\ is relatively low. Even at the lowest fluxes the source density is $<0.1$ arcmin$^{-2}$ (4.6\,kpc$^{-2}$), which is significantly lower than the surface density of background sources. The cumulative $\log N$--$\log S$ of the sources in \m\ is also significantly flatter than that of the background, so that the surface density contributions are equivalent at $f_X\simeq10^{-14}$\erg\,cm$^{-2}$ ($L_X\simeq10^{36}$\erg\ for objects in \m). When the SNRs and FSs are removed from the \m\ source LF, the function becomes even flatter. 

The number of HMXBs should correlate with the star formation rate \citep{grimm2003} and the number of LMXBs should correlate with the total stellar mass \citep{gil04}.  We adopt a star formation rate for \m\ of  $0.45\pm0.10$\,$M_{\odot}/{\rm yr}$ from \cite{verley2009} and a stellar mass of $4.5\times10^{9}$\,$M_{\odot}$ \citep{corbelli2003}. In order to scale from values determined for the Milky Way, we adopt a stellar mass of the Milky Way of $5.0\times10^{10}$\,$M_{\odot}$ \citep{hammer2007} and a range of star formation rates from $0.68-5.0$\,$M_{\odot}$ \citep{smith1978,robitaille2010}. Given the star formation rate and total stellar mass of \m\, we expect that HMXBs will dominate over LMXBs for luminosities larger than $L_X>10^{36}$\erg, but the expected number of HMXBs and LMXBs is roughly equal for luminosities larger than $L_X>10^{37}$\erg given the steeper slope of the HMXB LF.
The estimates of the XRB cumulative LFs in Figure~\ref{f7} suggest a rather flat slope.  \cite{grimm2002} predict a slope of 1.6 for HMXBs and 1.2 for LMXBs based on the populations in the Milky Way. Our estimate of the slope is 1.5 based on the soft band luminosity function, but the uncertainties are large  given the large uncertainty in the AGN model. 

There are several methods for estimating the number of HMXBs, LMXBs, and total XRBs expected in \m.  We have examined the predictions from three methods: (1) simply scaling from the observed population in the MW \citep{grimm2002}, (2) population synthesis modeling for HMXBs \cite{dalton1995}, and (3) ``universal'' LFs for HMXBs \citep{grimm2003} and LMXBs \citep{gil04} based on the analysis of \chandra\/ data of nearby galaxies. We compare the expected number of total XRBs (where total is the sum of the HMXBs and LMXBs) to the number of XRBs derived from the \chase\/ survey  LFs in three luminosity ranges ($L_X>10^{38},10^{37}$, \& $10^{36}$\erg), except for the \cite{dalton1995} results which only predict the number of HMXBs.
The energy band from 2.0-10.0~keV has been used traditionally for the luminosity of XRBs, so we converted our 2.0-8.0~keV band fluxes to the broader band assuming a disk blackbody plus power-law model appropriate for XRBs.

We estimate that in the \chase\/ source catalog there are 2/7/16 XRBs more luminous than $L_X>10^{38}/10^{37}/10^{36}$\erg\/ respectively. We scaled the expected numbers by the star formation rate and stellar mass of \m\/ and also by a factor of 0.7 to account for the fact that the \chase\/ survey only covers 70\% of the area of the galaxy within the D$_{25}$ isophote.  Using the LFs for HMXBs and LMXBs in the Milky Way derived by \cite{grimm2002}, we would expect (0.6-1.0)/(2.4-4.2)/(6.4-14.3) XRBs more luminous than $L_X>10^{38}/10^{37}/10^{36}$\erg\/ respectively, where the range of values is due to the assumed range of star formation rates for the Milky Way. 
The \cite{dalton1995} models predict 1\,--\,6 HMXBs with $L_X>10^{37}$\erg\/ and 4\,--\,26 HMXBs with $L_X>10^{36}$\erg\/ (they made no predictions for  $L_X>10^{38}$\erg\/).  Finally, the universal LFs of \cite{grimm2003} and \cite{gil04} predict a total number of XRBs more luminous than $L_X>10^{38}/10^{37}/10^{36}$\erg\/ of 2.3/11.4/39.0 respectively. 

Our estimates and the model predictions agree to within a factor of $\sim$3.  The predictions based on the Milky Way populations are lower than our estimates for \m\/, while the predictions from the universal LFs are higher than our estimates for \m. Given the rather crude scaling by star formation rate and total stellar mass, the inherent uncertainty in the universal LFs (estimated to be 50\%), and the partial coverage of \m\/ by the \chase\/ survey, agreement to within a factor of $\sim$3 is not surprising. Agreement to this level indicates that the processes which control the formation and evolution of XRBs in \m\/ are not radically different from the Milky Way, M31, and other nearby galaxies.

\section{Summary and Conclusions}\label{sec-summary}
We have presented the final version of the X-ray point source catalog in the framework of the \chase\ project. \chase\ is the deepest and spatially best resolved X-ray survey of any galaxy so far. A total of 662 sources were detected of which at least 100 could be identified to be located in \m. The faintest detected point source in the 0.35\,--\,8.0\,keV energy band (\#501) has an unabsorbed energy (photon) flux of $2.79\times10^{-16}$\erg\ cm$^{-2}$ ($1.42\times10^{-7}$\,photons s$^{-1}$ cm$^{-2}$) if we adopt our standard model with $\Gamma=1.9$ and an \nh\ of $6\times10^{20}$\,cm$^{-2}$. If this source is located in \m, its luminosity would be $2.2\times10^{34}$\erg. By far the brightest source in \m\ is the nucleus (\#318) with a total unabsorbed energy (photon) flux of $1.5\times10^{-11}$\erg\ cm$^{-2}$ ($6.2\times10^{-3}$\,photons s$^{-1}$ cm$^{-2}$), which translates into a luminosity of about $1.2\times10^{39}$\erg, making this object the brightest single X-ray source in the Local Group.

Neither of the supersoft X-ray sources reported by PMH07 and MPH06 were detected nor did we detect new candidates. This, however, is not surprising as the sensitivity of the front-illuminated ACIS-I chips is insufficient for such studies. Among the 38 \chase\ sources which are time-variable, 35 (7) are variable on short (long) time scales, while 4 show both types of variability.

Based on optical follow-up spectroscopy, the analysis of multi-wavelength data, and the cross correlation with other catalogs, we were able to identify counterparts for 183 of the 662 X-ray sources. Most sources, which appear to be located within \m, seem to be well aligned with the spiral arms of this galaxy. By means of hardness ratio diagrams we could identify two distinct regions, one covered by SNRs and foreground stars and one region which is populated by the bulk of (yet unidentified) sources. The majority of unidentified sources are likely background galaxies/AGN.

We were able to perform detailed spectral fits to the 15 sources with the largest number of net counts (11 in this paper and 4 in previous papers).  The detailed spectral fits confirmed or supported the classification of 5 of the 11 sources as XRBs when the variability is also considered. Three of the 11 sources are confirmed AGN based on the X-ray spectra, X-ray variability, and measured redshift, while 3 of the 11 sources are most likely AGN based on the X-ray spectra and variability.

We presented the differential  $\log N$--$\log S$ of all the sources in the \chase\ in the 0.5--2.0~keV and 2.0--8.0~keV bands.  Taking advantage of the 183 source classifications, we were able to create differential $\log N$--$\log S$ functions for the foreground stars and SNRs in \m.  We modeled the contribution of background AGN and subtracted it from the total  $\log N$--$\log S$ along with the foreground stars and SNRs to generate a $\log N$--$\log S$ of the XRBs and unidentified sources which could be in \m. The largest uncertainty in creating a  $\log N$--$\log S$ of the XRBs and unclassified sources in \m\ is the model of the AGN contribution.  We therefore explored the sensitivity of our result for a range of AGN models.  We then created cumulative LFs for the SNRs and XRBs and unidentified sources in \m. The slope of the LF for XRBs is closer to 1.6, the expected value for a dominant HMXB population.  Additionally, the number of candidate XRBs above  $10^{35}$ erg s$^{-1}$, $10^{36}$ erg s$^{-1}$, and $10^{37}$ erg s$^{-1}$ agrees to within a factor of three with the number of XRBs in the Milky Way after scaling by stellar mass and star formation rate, the number of HMXBs predicted by population synthesis models, and the number derived from universal luminosity functions based on XRBs in other nearby galaxies.

\acknowledgments
Support for this work was provided by the National Aeronautics and Space Administration through \chandra\ Award Number G06-7073A issued by the \chandra\ X-ray Observatory Center, which is operated by the Smithsonian Astrophysical Observatory and on behalf of the National Aeronautics Space and Administration under contract NAS8-03060. 
RT, PPP, TJG, and RJE  acknowledge support under NASA contract NAS8-03060. This work has made use of SAOImage DS9, developed by the Smithsonian Astrophysical Observatory \citep{joye03}, the {\tt XSPEC} spectral fitting package (Arnaud 1996), the FUNTOOLS utilities package, the HEASARC FTOOLS package, the CIAO (\chandra{} Interactive Analysis of Observations) package, and {\tt ACIS Extract}, the source extraction and characterization tool developed and maintained by Pat Broos. Sincere thanks to Pat Broos for answering our questions regarding \ace\ and for maintaining this important tool and to David Thilker for kindly providing the \hi\ map of \m. We also acknowledge the heroic effort of the anonymous referee for a careful reading of the manuscript and several comments that helped to improve the paper.

\appendix
\section{APPENDIX}\label{app}
This section provides all tables which contain the basic \chase\ catalog data. Table~\ref{sourcelist} lists basic properties for each source, such as source position, its positional uncertainty, exposure time, and size of the source and background region. In Table~\ref{mergedpns} we provide the merged $pns$ values for each source in eight different energy bands, while Tables~\ref{netcts} and \ref{mergedphotfl} list the corresponding net counts and photon fluxes in these bands. 

In Table~\ref{bestfit} we provide the results of the spectral analysis of sources with 8 or more spectral groups. We list the best-fit model together with the best-fit parameters, as well as deabsorbed X-ray luminosities in the 0.35\,--\,2.0\,keV and 0.35\,--\,8.0\,keV energy bands. As some of the source spectra have more than 2000 counts, we fit those sources with more complex models. These results are listed in Table~\ref{11fitspect}.  

The results of the cross correlation of the \chase\ data with other multi-wavelength spectrophotometric data are given in Table~\ref{crossref} together with $H\alpha$ surface brightnesses and results from our variability analysis of the X-ray sources. Finally, Table~\ref{crosss} provides the cross-identifications of the SSSs between the \chase\ catalog and the catalogs from PMH04 and MPH06.

\begin{deluxetable}{ccccccccccc}
\tabletypesize{\footnotesize}
\rotate
\tablewidth{0pc} 
\tablecolumns{10}
\tablecaption{\label{sourcelist} \chase\ source list.}
\tablehead{
\colhead{Source} & 
\colhead{Source ID} & 
\colhead{RA (J2000)} & 
\colhead{Dec (J2000)} & 
\colhead{Positional} & 
\colhead{No. of} & 
\colhead{Total ex-} & 
\colhead{Total exp. map} & 
\colhead{$R_{src}$} & 
\colhead{$R_{bkg}$} &
\colhead{$\theta$} \\
\colhead{No.}    &
\colhead{}              & 
\colhead{[\degr]}    & 
\colhead{[\degr]}     & 
\colhead{Error [\arcsec]}      & 
\colhead{obs.}   & 
\colhead{posure [s]}           & 
\colhead{value [$s\ cm^2$]}  & 
\multicolumn{2}{c}{[sky pixel]} &
\colhead{[$\arcmin$]} \\
\colhead{(1)} & 
\colhead{(2)} & 
\colhead{(3)} & 
\colhead{(4)} & 
\colhead{(5)} & 
\colhead{(6)} & 
\colhead{(7)} & 
\colhead{(8)} & 
\colhead{(9)} &
\colhead{(10)} & 
\colhead{(11)} 
}
\startdata
  1  &  013224.55+303322.3  &  23.102333  &  30.55622  &  2.34  &   1  &   96659  &  1.316e+07  &  24.3  &  66.1 & 11.6  \\
  2  &  013227.18+303521.0  &  23.113250  &  30.58919  &  1.17  &   1  &   96659  &  1.795e+07  &  18.3  &  56.7 & 10.1  \\
  3  &  013229.23+303618.6  &  23.121792  &  30.60517  &  0.74  &   1  &   96659  &  1.797e+07  &  15.8  &  44.8 & 9.2   \\
  4  &  013229.31+304514.0  &  23.122138  &  30.75389  &  1.40  &   1  &   96662  &  1.439e+07  &  17.5  &  50.1 & 9.8   \\
  5  &  013230.28+303548.1  &  23.126172  &  30.59671  &  0.97  &   1  &   96659  &  1.876e+07  &  15.6  &  43.4 & 9.3   \\
  6  &  013230.54+303618.0  &  23.127250  &  30.60500  &  1.44  &   1  &   96659  &  1.885e+07  &  15.4  &  42.0 & 9.0   \\
  7  &  013232.83+304027.9  &  23.136833  &  30.67442  &  1.43  &   2  &  187395  &  2.958e+07  &  11.9  &  33.2 & 7.7   \\
  8  &  013233.76+304726.7  &  23.140678  &  30.79077  &  1.62  &   2  &  187398  &  3.506e+07  &  19.7  &  67.6 & 10.3  \\
\enddata \\
\tablenotetext{\mbox{}}{(2): The source ID also contains the source coordinates (J2000.0). (5): The positional uncertainty is a simple error circle (in \arcsec) around RA and Dec. (7): Total exposure in all observations. (8): Sum of mean exposure map values in source region for all observations. (9) and (10): Average radius of the source and background extraction region (1 sky pixel = 0\farcs492). (11): Average off-axis angle.}
\end{deluxetable}

\begin{deluxetable}{ccccccccc}
\tabletypesize{\footnotesize}
\rotate
\tablewidth{0pc} 
\tablecolumns{9}
\tablecaption{\label{mergedpns} Merged $pns$ values for different energy bands$^{\dagger}$.}
\tablehead{
\colhead{Source No}   & 
\colhead{$pns$[1]} & 
\colhead{$pns$[2]} & 
\colhead{$pns$[3]} & 
\colhead{$pns$[4]} & 
\colhead{$pns$[5]} & 
\colhead{$pns$[6]} & 
\colhead{$pns$[7]} & 
\colhead{$pns$[8]} \\
\colhead{}            &
\colhead{[0.5\,--\,8.0keV]}  & 
\colhead{[0.5\,--\,2.0keV]}  & 
\colhead{[2.0\,--\,8.0keV]}  & 
\colhead{[0.35\,--\,8.0keV]} & 
\colhead{[0.35\,--\,1.1keV]} & 
\colhead{[1.1\,--\,2.6keV]}  & 
\colhead{[2.6\,--\,8.0keV]}  & 
\colhead{[0.35\,--\,2.0keV]}
}
\startdata
  1 & 	$<$1.0e-10 & 	1.8e-10 & 	1.9e-04 & 	$<$1.0e-10 & 	8.7e-02 & 	$<$1.0e-10 & 	2.3e-03 & 	2.7e-10 \\
  2 & 	1.7e-06 & 	7.4e-06 & 	5.6e-03 & 	1.7e-06 & 	4.0e-02 & 	6.0e-05 & 	1.4e-02 & 	7.4e-06 \\
  3 & 	$<$1.0e-10 & 	$<$1.0e-10 & 	$<$1.0e-10 & 	$<$1.0e-10 & 	3.2e-10 & 	$<$1.0e-10 & 	4.9e-10 & 	$<$1.0e-10 \\
  4 & 	1.4e-09 & 	$<$1.0e-10 & 	3.6e-02 & 	1.4e-09 & 	1.7e-04 & 	1.2e-10 & 	2.6e-01 & 	$<$1.0e-10 \\
  5 & 	$<$1.0e-10 & 	$<$1.0e-10 & 	4.8e-07 & 	$<$1.0e-10 & 	2.4e-03 & 	$<$1.0e-10 & 	2.0e-05 & 	$<$1.0e-10 \\
  6 & 	$<$1.0e-10 & 	$<$1.0e-10 & 	4.2e-01 & 	$<$1.0e-10 & 	$<$1.0e-10 & 	$<$1.0e-10 & 	8.0e-01 & 	$<$1.0e-10 \\
  7 & 	1.7e-06 & 	2.4e-06 & 	4.8e-03 & 	2.2e-06 & 	3.5e-01 & 	1.5e-05 & 	5.6e-03 & 	4.0e-06 \\
  8 & 	2.8e-04 & 	$<$1.0e-10 & 	8.2e-01 & 	3.4e-04 & 	$<$1.0e-10 & 	5.7e-02 & 	8.6e-01 & 	$<$1.0e-10 \\
\enddata  
\tablenotetext{\mbox{}}{$^{\dagger}$The $pns$ value is the Poission probability of not being a source. $pns$ values smaller than 1.0e-10 have been replaced by $<$1.0e-10. All these cases are highly significant detections.}
\end{deluxetable}

\begin{deluxetable}{ccccccccc}
\tabletypesize{\footnotesize}
\rotate
\tablewidth{0pc} 
\tablecolumns{9}
\tablecaption{\label{netcts} Total net counts in different energy bands$^{\dagger}$.}
\tablehead{
\colhead{Source No}   & 
\colhead{$net\_cnts$[1]} & 
\colhead{$net\_cnts$[2]} & 
\colhead{$net\_cnts$[3]} & 
\colhead{$net\_cnts$[4]} & 
\colhead{$net\_cnts$[5]} & 
\colhead{$net\_cnts$[6]} & 
\colhead{$net\_cnts$[7]} & 
\colhead{$net\_cnts$[8]} \\
\colhead{}            &
\colhead{[0.5\,--\,8.0keV]}  & 
\colhead{[0.5\,--\,2.0keV]}  & 
\colhead{[2.0\,--\,8.0keV]}  & 
\colhead{[0.35\,--\,8.0keV]} & 
\colhead{[0.35\,--\,1.1keV]} & 
\colhead{[1.1\,--\,2.6keV]}  & 
\colhead{[2.6\,--\,8.0keV]}  & 
\colhead{[0.35\,--\,2.0keV]}
}
\startdata
  1 & 	    47.6$^{+	     9.9}_{-	     8.8}$ &	    27.6$^{+	     7.1}_{-	     6.0}$ &	    20.0$^{+	     7.6}_{-	     6.5}$ &	    47.5$^{+	     9.9}_{-	     8.8}$ &	     3.9$^{+	     4.1}_{-	     2.9}$ &	    29.0$^{+	     7.1}_{-	     6.0}$ &	    14.6$^{+	     6.9}_{-	     5.8}$ &	    27.4$^{+	     7.1}_{-	     6.0}$ \\
  2 & 	    28.1$^{+	     8.3}_{-	     7.2}$ &	    15.5$^{+	     5.8}_{-	     4.6}$ &	    12.6$^{+	     6.6}_{-	     5.5}$ &	    28.1$^{+	     8.3}_{-	     7.2}$ &	     3.9$^{+	     3.7}_{-	     2.4}$ &	    14.3$^{+	     5.8}_{-	     4.7}$ &	     9.8$^{+	     6.0}_{-	     4.9}$ &	    15.5$^{+	     5.8}_{-	     4.6}$ \\
  3 & 	   101.1$^{+	    12.0}_{-	    10.9}$ &	    57.1$^{+	     8.9}_{-	     7.8}$ &	    44.0$^{+	     8.6}_{-	     7.5}$ &	   102.1$^{+	    12.0}_{-	    10.9}$ &	    17.7$^{+	     5.5}_{-	     4.3}$ &	    53.5$^{+	     8.7}_{-	     7.6}$ &	    30.8$^{+	     7.7}_{-	     6.5}$ &	    58.1$^{+	     9.0}_{-	     7.9}$ \\
  4 & 	    32.8$^{+	     8.1}_{-	     7.0}$ &	    25.1$^{+	     6.4}_{-	     5.3}$ &	     7.6$^{+	     5.6}_{-	     4.5}$ &	    32.8$^{+	     8.1}_{-	     7.0}$ &	     7.9$^{+	     4.2}_{-	     3.0}$ &	    22.1$^{+	     6.2}_{-	     5.0}$ &	     2.8$^{+	     4.9}_{-	     3.7}$ &	    25.1$^{+	     6.4}_{-	     5.3}$ \\
  5 & 	    48.9$^{+	     9.0}_{-	     7.9}$ &	    26.8$^{+	     6.5}_{-	     5.4}$ &	    22.1$^{+	     6.8}_{-	     5.7}$ &	    49.5$^{+	     9.0}_{-	     7.9}$ &	     5.2$^{+	     3.7}_{-	     2.4}$ &	    27.2$^{+	     6.6}_{-	     5.5}$ &	    17.1$^{+	     6.2}_{-	     5.1}$ &	    27.4$^{+	     6.6}_{-	     5.5}$ \\
  6 & 	    74.4$^{+	    10.8}_{-	     9.7}$ &	    73.1$^{+	     9.8}_{-	     8.7}$ &	     1.3$^{+	     5.2}_{-	     4.1}$ &	    76.1$^{+	    10.9}_{-	     9.8}$ &	    35.6$^{+	     7.3}_{-	     6.2}$ &	    43.0$^{+	     7.8}_{-	     6.7}$ &	    -2.5$^{+	     4.6}_{-	     3.4}$ &	    74.8$^{+	     9.9}_{-	     8.9}$ \\
\enddata  
\tablenotetext{\mbox{}}{$^{\dagger}$Uncertainties are given on a 90\% confidence level.}
\end{deluxetable}
\begin{deluxetable}{crrrrrrrr}
\tabletypesize{\footnotesize}
\rotate
\tablewidth{0pc} 
\tablecolumns{9}
\tablecaption{\label{mergedphotfl} Merged photon flux ($photons/cm^2/s$) in different energy bands.}
\tablehead{
\colhead{Source No}   & 
\colhead{$flux$[1]} & 
\colhead{$flux$[2]} & 
\colhead{$flux$[3]} & 
\colhead{$flux$[4]} & 
\colhead{$flux$[5]} & 
\colhead{$flux$[6]} & 
\colhead{$flux$[7]} & 
\colhead{$flux$[8]} \\
\colhead{}            &
\colhead{[0.5\,--\,8.0keV]}  & 
\colhead{[0.5\,--\,2.0keV]}  & 
\colhead{[2.0\,--\,8.0keV]}  & 
\colhead{[0.35\,--\,8.0keV]} & 
\colhead{[0.35\,--\,1.1keV]} & 
\colhead{[1.1\,--\,2.6keV]}  & 
\colhead{[2.6\,--\,8.0keV]}  & 
\colhead{[0.35\,--\,2.0keV]}
}
\startdata
  1 & 	4.10{\rm e}$-$06 & 	1.82{\rm e}$-$06 & 	1.87{\rm e}$-$06 & 	4.16{\rm e}$-$06 & 	7.65{\rm e}$-$07 & 	1.61{\rm e}$-$06 & 	1.40{\rm e}$-$06 & 	1.98{\rm e}$-$06 \\
  2 & 	1.54{\rm e}$-$06 & 	6.39{\rm e}$-$07 & 	7.53{\rm e}$-$07 & 	1.57{\rm e}$-$06 & 	4.57{\rm e}$-$07 & 	5.09{\rm e}$-$07 & 	6.03{\rm e}$-$07 & 	7.00{\rm e}$-$07 \\
  3 & 	5.81{\rm e}$-$06 & 	2.46{\rm e}$-$06 & 	2.76{\rm e}$-$06 & 	5.98{\rm e}$-$06 & 	2.11{\rm e}$-$06 & 	2.00{\rm e}$-$06 & 	1.98{\rm e}$-$06 & 	2.74{\rm e}$-$06 \\
  4 & 	2.45{\rm e}$-$06 & 	1.42{\rm e}$-$06 & 	6.21{\rm e}$-$07 & 	2.49{\rm e}$-$06 & 	1.25{\rm e}$-$06 & 	1.08{\rm e}$-$06 & 	2.29{\rm e}$-$07 & 	1.56{\rm e}$-$06 \\
  5 & 	2.63{\rm e}$-$06 & 	1.08{\rm e}$-$06 & 	1.30{\rm e}$-$06 & 	2.71{\rm e}$-$06 & 	5.81{\rm e}$-$07 & 	9.49{\rm e}$-$07 & 	1.03{\rm e}$-$06 & 	1.21{\rm e}$-$06 \\
  6 & 	3.93{\rm e}$-$06 & 	2.89{\rm e}$-$06 & 	7.45{\rm e}$-$08 & 	4.09{\rm e}$-$06 & 	3.87{\rm e}$-$06 & 	1.48{\rm e}$-$06 & $-$1.46{\rm e}$-$07 & 	3.24{\rm e}$-$06 \\
  7 & 	8.71{\rm e}$-$07 & 	3.37{\rm e}$-$07 & 	4.58{\rm e}$-$07 & 	8.81{\rm e}$-$07 & 	6.30{\rm e}$-$08 & 	2.95{\rm e}$-$07 & 	4.26{\rm e}$-$07 & 	3.64{\rm e}$-$07 \\
  8 & 	7.41{\rm e}$-$07 & 	6.63{\rm e}$-$07 & $-$1.47{\rm e}$-$07 & 	7.51{\rm e}$-$07 & 	1.60{\rm e}$-$06 & 	1.02{\rm e}$-$07 & $-$1.70{\rm e}$-$07 & 	7.23{\rm e}$-$07 \\
\enddata  
\end{deluxetable}
\begin{deluxetable}{cccccccccrr}
\tabletypesize{\scriptsize}
\rotate
\tablewidth{0pc} 
\tablecolumns{11}
\tablecaption{\label{bestfit} Best-fit results for sources with 8 or more spectral groups.}
\tablehead{
\colhead{No.}   & 
\colhead{Source ID}   & 
\colhead{Best-fit$^a$} & 
\colhead{$N_H^b$} & 
\colhead{$\Gamma$} & 
\colhead{$kT$} & 
\colhead{Norm$^c$} & 
\colhead{DOF$^d$} & 
\colhead{$\chi^2_{red}$} &
\colhead{$L_X$ (0.35\,--\,2.0keV)$^e$}  &
\colhead{$L_X$ (0.35\,--\,8.0keV)$^f$}  \\
\colhead{}  &
\colhead{}  &
\colhead{model}  & 
\colhead{($10^{22}$\,cm$^{-2}$)}  & 
\colhead{}  & 
\colhead{(keV)} & 
\colhead{} & 
\colhead{}  & 
\colhead{}  & 
\multicolumn{2}{c}{($10^{35}$\,erg s$^{-1}$)} 
}
\startdata
3 & 013229.23+303618.6 & pow &	0.01$^{+0.00}_{-0.00}$ & 	1.37$^{+0.00}_{-0.00}$ & 	-- & 	2.54e-06$^{+0.00e+00}_{-0.00e+00}$ & 	   5 &	0.96 & 	4.51$^{+0.37}_{-4.43}$ & 	15.61$^{+2.03}_{-15.53}$ \\
6 & 013230.54+303618.0 & vapec &	0.01$^{+0.00}_{-0.00}$ & 	-- & 	1.05$^{+0.00}_{-0.00}$ & 	8.40e-06$^{+0.00e+00}_{-0.00e+00}$ & 	   5 &	0.94 & 	4.84$^{+0.38}_{-2.89}$ & 	5.62$^{+0.47}_{-2.36}$ \\
10 & 013236.84+303229.0 & pow &	0.45$^{+0.23}_{-0.20}$ & 	1.07$^{+0.14}_{-0.13}$ & 	-- & 	1.73e-05$^{+3.28e-06}_{-2.70e-06}$ & 	  76 &	1.64 & 	31.61$^{+1.53}_{-1.49}$ & 	150.34$^{+8.29}_{-9.09}$ \\
13 & 013241.32+303217.7 & pow &	0.50$^{+1.91}_{-0.50}$ & 	0.83$^{+0.84}_{-0.58}$ & 	-- & 	5.33e-07$^{+9.70e-07}_{-2.83e-07}$ & 	   7 &	1.51 & 	1.01$^{+0.19}_{-0.37}$ & 	6.31$^{+1.67}_{-5.89}$ \\
14 & 013242.07+303329.0 & pow &	0.50$^{+1.42}_{-0.50}$ & 	2.10$^{+1.71}_{-0.91}$ & 	-- & 	1.43e-06$^{+4.25e-06}_{-8.29e-07}$ & 	   5 &	1.15 & 	2.58$^{+0.43}_{-1.23}$ & 	4.78$^{+1.11}_{-3.39}$ \\
15 & 013242.46+304815.0 & pow &	0.19$^{+0.76}_{-0.19}$ & 	1.82$^{+0.81}_{-0.49}$ & 	-- & 	1.56e-06$^{+1.62e-06}_{-5.70e-07}$ & 	   5 &	0.95 & 	2.76$^{+0.53}_{-0.73}$ & 	6.31$^{+0.71}_{-1.89}$ \\
16 & 013243.41+303506.2 & pow &	0.34$^{+0.00}_{-0.00}$ & 	1.67$^{+0.00}_{-0.00}$ & 	-- & 	6.17e-06$^{+0.00e+00}_{-0.00e+00}$ & 	  21 &	2.04 & 	10.87$^{+0.75}_{-0.99}$ & 	28.20$^{+2.94}_{-2.80}$ \\
22 & 013247.71+304711.1 & vapec &	0.56$^{+0.00}_{-0.00}$ & 	-- & 	0.77$^{+0.00}_{-0.00}$ & 	5.61e-06$^{+0.00e+00}_{-0.00e+00}$ & 	   7 &	1.62 & 	3.57$^{+0.24}_{-1.05}$ & 	3.83$^{+0.31}_{-1.02}$ \\
\enddata  
\tablenotetext{a}{The best-fit model is either a {\tt powerlaw} or a {\tt vapec} model, depending which one has the lower $\chi^2_{red}$. $^b$This is the column density internal to \m. The Galactic $N_H$ was fixed at $0.06\times10^{22}$\,cm$^{-2}$ (Dickey \& Lockman 1990). $^c$Normalization constant of the fit. For the {\tt powerlaw} and the {\tt vapec} model, the units are photons keV$^{-1}$\,cm$^{-2}$\,s$^{-1}$ at 1\,keV and cm$^{-5}$, respectively. $^d$Degrees of freedom. $^e$and $^f$absorption-corrected X-ray luminosities assume a distance to \m\ of $D=817$\,kpc (Freedman et al. 2001). All uncertainties are given at a 90\% confidence level. 
}
\end{deluxetable}
\begin{deluxetable}{cccccccrccc}
\tabletypesize{\scriptsize}
\rotate
\tablewidth{0pc} 
\tablecolumns{9}
\tablecaption{\label{11fitspect} Refined Spectral Fit Results for Bright Sources}
\tablehead{
\colhead{} & 
\colhead{} &
\colhead{} &
\colhead{} & 
\multicolumn{2}{c}{{\tt diskbb}} &
\multicolumn{2}{c}{{\tt pow}} &
\colhead{ } &
\multicolumn{2}{c}{deabsorbed $L_{\rm X}$$^d$} \\
\colhead{Source ID}   & 
\colhead{Source} &
\colhead{Total} &
\colhead{$N_H^a$}   & 
\colhead{$kT$\,(keV)} & 
\colhead{Norm$^b$} & 
\colhead{$\Gamma$} & 
\colhead{Norm$^c$} & 
\colhead{$\chi^2_{\bf red}$/DOF} &
\colhead{(0.35\,--\,2.0keV)} & 
\colhead{(0.35\,--\,8.0keV)} \\
\colhead{ }  &
\colhead{No.}  &
\colhead{Counts}  &
\colhead{($10^{22}$\,cm$^{-2}$)}   & 
\colhead{ }  &
\colhead{ }  &
\colhead{ }  &
\colhead{($\times 10^{-4}$)}  &
\colhead{ } &
\multicolumn{2}{c}{{(erg s$^{-1}$)}} \\ 
}
\startdata
013350.89+303936.6 & 318 & 163938 & 0.28$^{+0.06}_{-0.05}$ & 0.95$^{+0.04}_{-0.04}$ & 60.5$^{+3.4}_{-3.8}$   & 1.20$^{+0.29}_{-0.40}$ & 5.61$^{+4.40}_{-3.11}$ & 1.02/237 & (5.43$^{+0.02}_{-0.12}$)$\times 10^{38}$ & (1.18$^{+0.04}_{-0.24}$)$\times 10^{39}$ \\
013328.69+302723.6 & 180 & 47082  & 0.00$^{+0.01}_{-0.00}$ & 1.50$^{+0.36}_{-0.26}$ & 6.48$^{+2.13}_{-1.54}$ & 1.37$^{+0.10}_{-0.06}$ & 2.28$^{+0.16}_{-0.16}$ & 1.35/241 & (6.54$^{+0.05}_{-0.62}$)$\times 10^{37}$ & (1.96$^{+0.02}_{-0.16}$)$\times 10^{38}$ \\
013324.40+304402.4 & 158 & 33863  & 0.22$^{+0.06}_{-0.06}$ & 1.21$^{+0.24}_{-0.16}$ & 7.00$^{+2.52}_{-1.82}$ & 1.81$^{+0.21}_{-0.18}$ & 1.11$^{+0.21}_{-0.25}$ & 1.21/239 & (3.51$^{+0.03}_{-0.33}$)$\times 10^{37}$ & (7.51$^{+0.09}_{-0.98}$)$\times 10^{37}$ \\
013451.85+302909.7 & 612 & 26303  & 0.29$^{+0.14}_{-0.09}$ & 2.62$^{+0.39}_{-1.07}$ & 3.63$^{+1.30}_{-1.15}$ & 1.94$^{+0.76}_{-0.44}$ & 3.43$^{+0.63}_{-0.47}$ & 0.99/236 & (1.03$^{+0.13}_{-0.03}$)$\times 10^{38}$ & (2.59$^{+0.59}_{-0.17}$)$\times 10^{38}$ \\
013315.16+305318.2 & 113 & 23117  & 0.53$^{+0.11}_{-0.11}$ & 0.92$^{+0.07}_{-0.07}$ & 22.6$^{+4.4}_{-3.4}$   & 2.51$^{+0.34}_{-0.34}$ & 2.74$^{+0.98}_{-1.05}$ & 1.13/239 & (1.17$^{+0.01}_{-0.06}$)$\times 10^{38}$ & (1.79$^{+0.03}_{-0.14}$)$\times 10^{38}$ \\
013425.80+305518.1 & 492 & 9064   & 0.19$^{+0.04}_{-0.04}$ & --                     & --                     & 1.87$^{+0.05}_{-0.05}$ & 0.95$^{+0.05}_{-0.05}$ & 1.00/238 & (2.08$^{+0.05}_{-0.07}$)$\times 10^{37}$ & (4.09$^{+0.10}_{-0.10}$)$\times 10^{37}$ \\
013253.89+303311.8 &  37 & 4891   & 0.17$^{+0.06}_{-0.05}$ & --                     & --                     & 1.95$^{+0.08}_{-0.08}$ & 0.42$^{+0.03}_{-0.03}$ & 0.91/167 & (9.18$^{+0.39}_{-0.34}$)$\times 10^{36}$ & (1.82$^{+0.07}_{-0.07}$)$\times 10^{37}$ \\
\enddata  
\tablenotetext{a}{ Column density internal to \m. The Galactic $N_H$ was fixed at $0.06\times10^{22}$\,cm$^{-2}$ \citep{dickey90}.  $^b$ Normalization constant for the disk blackbody model {\tt diskbb} in units of (km), where the normalization is $R_{in}(cos\theta)^{1/2}$ with $R_{in}$ being the inner disk radius and $\theta$ being the inclination angle (assuming a distance to M33 of $D=817$\,kpc).  $^c$ Normalization constant for the non-thermal model {\tt pow} in units of  photons keV$^{-1}$\,cm$^{-2}$\,s$^{-1}$ at 1\,keV. $^d$ Deabsorbed X-ray luminosities assume a distance to \m\ of 817\,kpc \citep{freed01}. All uncertainties are given at a 90\% confidence level.
}
\end{deluxetable}

\tabletypesize{\scriptsize}
\begin{deluxetable}{cccccccccccc}
\rotate
\tablewidth{0pc} 
\tablecolumns{11}
\tablecaption{\label{crossref} Cross-referenced source list.}
\tablehead{
\colhead{Source ID$^a$} & 
\colhead{Src.} & 
\colhead{FLC} & 
\colhead{XMM} & 
\colhead{G05} & 
\colhead{Extent} & 
\multicolumn{3}{c}{Variability} &
\colhead{$S_{H\alpha}$} & 
\colhead{Source type} & 
\colhead{Comments} \\
\colhead{} & 
\colhead{No.} & 
\colhead{No.} & 
\colhead{No.} & 
\colhead{No.} & 
\colhead{} & 
\colhead{$\eta$} & 
\colhead{$\xi$} & 
\colhead{$\frac{flux_{max}}{flux_{min}}$} & 
\colhead{{\tiny ($10^{-17}$\,erg\,$s^{-1}$\,cm$^{-2}$\,arcsec$^{-2}$)}} & 
\colhead{} & 
\colhead{} \\
\colhead{(1)} & 
\colhead{(2)} & 
\colhead{(3)} & 
\colhead{(4)} & 
\colhead{(5)} & 
\colhead{(6)} & 
\colhead{(7)} & 
\colhead{(8)} & 
\colhead{(9)} & 
\colhead{(10)} & 
\colhead{(11)} &
\colhead{(12)} 
} 
\startdata
013224.55+303322.3   &    1   &    1   &  30,28   &   --   &   0   &  -- 	        & -- 		  & --		        & -0.187	   & --                        	        & --                                    \\
013227.18+303521.0   &    2   &    2   &  --,--   &   --   &   0   &	-- 	        & -- 		  & --		        & 2.157	  	   & --  			        & --                                    \\    
013229.23+303618.6   &    3   &    3   &  35,35   &   --   &   0   &	-- 	        & -- 		  & --		        & 16.261	   & Galaxy,2MASS-3086,                 & --                                    \\ 
                     &        &        &          &        &       &	    	        &    		  &    		        &       	   & USNO-7682                          &                                       \\ 
013229.31+304514.0   &    4   &    4   &  37,36   &   --   &   0   &	-- 	        & -- 		  & --		        & -0.692	   & --  			        & pos. uncertain                        \\    
013230.28+303548.1   &    5   &    5   &  --,--   &   --   &   0   &	-- 	        & -- 		  & --		        & 1.456	  	   & non-stellar		        & --                                    \\    
013230.54+303618.0   &    6   &    6   &  39,--   &   --   &   0   &	-- 	        & -- 		  & --		        & 31.397	   & FS,2MASS-3043,                     & --                                    \\ 
                     &        &        &          &        &       &	    	        &    		  &    		        &       	   & USNO-7490                          &                                       \\
\enddata
\end{deluxetable}

\begin{table}
{\scriptsize
\caption{\label{crosss} Cross identifications of supersoft sources}
\begin{tabular}{cc|cc|cc}
\hline
\multicolumn{2}{c|}{PMH04} & \multicolumn{2}{c|}{MPH06} & \multicolumn{2}{c}{\chase} \\
Source ID & Comments & Source ID & Comments & Source ID & Comments \\
\hline
52 & 01:32:38.01 +30:39:02.5 & 48 & 01:32:37.91 +30:39:03.7 & -- & not detected, \\
   &                         &    &                         &    & at border of FOV \\
-- &                         & 86 & 01:32:59.11 +30:35:45.5 & -- & not detected     \\
   &                         &    & only in one ObsID       &    & \\
-- &                         & 92 & 01:33:01.94 +31:00:25.9 & -- & outside FOV \\
   &                         &    & only in one ObsID       &    & \\
\end{tabular}
}
\end{table}


\clearpage
\begin{thebibliography}{16}
\bibitem[Alexander et al.(2003)]{alex03} Alexander, D.~M., et al.\ 2003, \aj, 126, 539 
\bibitem[Broos et al.(2002)]{Broos02} Broos, P. S., Townsley, L. K., Getman, K., \& Bauer, F. E.\ 2002, ACIS Extract, An ACIS Point Source Extraction Package (University Park: Pennsylvania State Univ.) http://www.astro.psu.edu/xray/docs/TARA/ae\_users\_guide.html
\bibitem[Broos et al.(2010)]{Broos10} Broos, P.~S., Townsley, L.~K., Feigelson, E.~D., Getman, K.~V., Bauer, F.~E., \& Garmire, G.~P.\ 2010, \apj, 714, 1582 
\bibitem[Capaccioli et al.(1989)]{cappa89} Capaccioli, M., della Valle, M., Rosino, L., \& D'Onofrio, M.\ 1989, \aj, 97, 1622 
\bibitem[Cappelluti et al.(2009)]{cap09} Cappelluti, N., et al.\ 2009, \aap, 497, 635 
\bibitem[Corbelli(2003)]{corbelli2003} Corbelli, E.\ 2003, \mnras, 342, 199 
\bibitem[Cutri et al.(2003)]{cutri03} Cutri, R.~M., et al.\ 2003, The IRSA 2MASS All-Sky Point Source Catalog, NASA/IPAC Infrared Science Archive.~http://irsa.ipac.caltech.edu/applications/Gator/
\bibitem[Dalton \& Sarazin(1995)]{dalton1995} Dalton, W.~W., \& Sarazin, C.~L.\ 1995, \apj, 440, 280 
\bibitem[Dickey \& Lockman(1990)]{dickey90} Dickey, J.~M., \& Lockman, F.~J.\ 1990, \araa, 28, 215 
\bibitem[Dubus \& Rutledge(2002)]{dubus02} Dubus, G., \& Rutledge, R.~E.\ 2002, \mnras, 336, 901 
\bibitem[Dubus et al.(2004)]{dubus04} Dubus, G., Charles, P.~A., \& Long, K.~S.\ 2004, \aap, 425, 95 
\bibitem[Fabricant et al.(2008)]{fabric08} Fabricant, D.~G., Kurtz, M.~J., Geller, M.~J., Caldwell, N., Woods, D., \& Dell'Antonio, I.\ 2008, \pasp, 120, 1222 
\bibitem[{{Freedman} {et~al.}(2001){Freedman}, {Madore}, {Gibson}, {Ferrarese}, {Kelson}, {Sakai}, {Mould}, {Kennicutt}, {Ford}, {Graham}, {Huchra}, {Hughes}, {Illingworth}, {Macri}, \& {Stetson}}]{freed01}{Freedman}, W.~L., {Madore}, B.~F., {Gibson}, B.~K., {Ferrarese}, L., {Kelson}, D.~D., {Sakai}, S., {Mould}, J.~R., {Kennicutt}, R.~C., {Ford}, H.~C., {Graham}, J.~A., {Huchra}, J.~P., {Hughes}, S.~M.~G., {Illingworth}, G.~D., {Macri}, L.~M., \& {Stetson}, P.~B. 2001, \apj, 553, 47
\bibitem[Freeman et al.(2002)]{freeman02} Freeman, P.~E., Kashyap, V., Rosner, R., \& Lamb, D.~Q.\ 2002, \apjs, 138, 185
\bibitem[Gaetz et al.(2007)]{gaetz07} Gaetz, T.~J., et al. 2007, \apj, 663, 234 
\bibitem[Gehrels(1986)]{gehrels86} Gehrels, N.\ 1986, \apj, 303, 336 
\bibitem[Georgakakis et al.(2008)]{george08} Georgakakis, A., Nandra, K., Laird, E.~S., Aird, J., \& Trichas, M.\ 2008, \mnras, 388, 1205 
\bibitem[Gilfanov(2004)]{gil04} Gilfanov, M.\ 2004, \mnras, 349, 146 
\bibitem[{{Gordon} {et~al.}(1998){Gordon}, {Kirshner}, {Long}, {Blair}, {Duric}, \& {Smith}}]{gordon98}{Gordon}, S.~M., {Kirshner}, R.~P., {Long}, K.~S., {Blair}, W.~P., {Duric}, N.,  \& {Smith}, R.~C. 1998, \apjs, 117, 89
\bibitem[Grimm et al.(2002)]{grimm2002} Grimm, H.-J., Gilfanov, M., \& Sunyaev, R.\ 2002, \aap, 391, 923 
\bibitem[Grimm et al.(2003)]{grimm2003} Grimm, H.-J., Gilfanov, M., \& Sunyaev, R.\ 2003, \mnras, 339, 793 
\bibitem[Grimm et al.(2005)]{grimm05} Grimm, H.-J., McDowell, J., Zezas, A., Kim, D.-W., \& Fabbiano, G.\ 2005, \apjs, 161, 271 [G05]
\bibitem[Haberl \& Pietsch(2001)]{hapi01} Haberl, F., \& Pietsch, W.\ 2001, \aap, 373, 438 [HP01]
\bibitem[Hammer et al.(2007)]{hammer2007} Hammer, F., Puech, M., Chemin, L., Flores, H., \& Lehnert, M.~D.\ 2007, \apj, 662, 322 
\bibitem[Harrison et al.(2003)]{harri03} Harrison, F.~A., Eckart, M.~E., Mao, P.~H., Helfand, D.~J., \& Stern, D.\ 2003, \apj, 596, 944 
\bibitem[Hartman et al.(2006)]{hart06} Hartman, J.~D., Bersier, D., Stanek, K.~Z., Beaulieu, J.-P., Kaluzny, J., Marquette, J.-B., Stetson, P.~B., \& Schwarzenberg-Czerny, A.\ 2006, \mnras, 371, 1405
\bibitem[Joye \& Mandel(2003)]{joye03} Joye, W.~A., \& Mandel, E.\ 2003, Astronomical Data Analysis Software and Systems XII, 295, 489 
\bibitem[Kahabka et al.(1999)]{kaha99} Kahabka, P., Pietsch, W., Filipovi{\'c} , M.~D., \& Haberl, F.\ 1999, \aaps, 136, 81 
\bibitem[Kim et al.(2007)]{kim07} Kim, M., et al.\ 2007, \apjs, 169, 401 
\bibitem[Kong et al.(2002)]{kong02} Kong, A.~K.~H., Garcia, M.~R., Primini, F.~A., Murray, S.~S., Di Stefano, R., \& McClintock, J.~E.\ 2002, \apj, 577, 738 
\bibitem[Kuntz et al.(2011)]{kuntzea11} Kuntz, K., et al. \ 2011, in prep.
\bibitem[Kuntz \& Snowden(2010)]{kuntz10} Kuntz, K.~D., \& Snowden, S.~L.\ 2010, \apjs, 188, 46 
\bibitem[La Parola et al.(2003)]{lapa03} La Parola, V., Damiani, F., Fabbiano, G., \& Peres, G.\ 2003, \apj, 583, 758 
\bibitem[Long et al.(1981)]{long81} Long, K.~S., Dodorico, S., Charles, P.~A., \& Dopita, M.~A.\ 1981, \apjl, 246, L61 
\bibitem[Long et al.(1996)]{long96} Long, K.~S., Charles, P.~A., Blair, W.~P., \& Gordon, S.~M.\ 1996, \apj, 466, 750 
\bibitem[Long et al.(2010)]{long10} Long, K. S., et al. \ 2010, \apjs,
  187, 495 [L10]
\bibitem[Markert \& Rallis(1983)]{mara83} Markert, T.~H., \& Rallis, A.~D.\ 1983, \apj, 275, 571 
\bibitem[Massey et al.(2006)]{mass06} Massey, P., Olsen, K.~A.~G., Hodge, P.~W., Strong, S.~B., Jacoby, G.~H., Schlingman, W., \& Smith, R.~C.\ 2006, \aj, 131, 2478
\bibitem[McNeil \& Winkler(2006)]{mcneil06} McNeil, E.~K. \& Winkler, P.~F. 2006, in Bull.\ AAS 39, 80
\bibitem[Misanovic et al.(2006)]{misa06} Misanovic, Z., Pietsch, W., Haberl, F., Ehle, M., Hatzidimitriou, D., \& Trinchieri, G.\ 2006, \aap, 448, 1247 [MPH06]
\bibitem[Monet et al.(2003)]{monet03} Monet, D.~G., et al.\ 2003, \aj, 125, 984
\bibitem[Nandra et al.(2005)]{nandra05} Nandra, K., et al.\ 2005, \mnras, 356, 568 
\bibitem[Newton(1980)]{newton1980} Newton, K.\ 1980, \mnras, 190, 689 
\bibitem[Orosz et al.(2007)]{oro07} Orosz, J.-A., et al.\ 2007, \nat, 449, 872
\bibitem[Park et al.(2006)]{park2006} Park, T., Kashyap, V.~L., Siemiginowska, A.,  van Dyk, D.~A., Zezas, A., Heinke, C., \& Wargelin, B.~J.\ 2006, \apj, 652, 610
\bibitem[Pence et al.(2001)]{pence01} Pence, W.~D., Snowden, S.~L., Mukai, K., \& Kuntz, K.~D.\ 2001, \apj, 561, 189 
\bibitem[{{Pietsch} {et~al.}(2004){pietsch04}, {Misanovic}, {Haberl},  {Hatzidimitriou}, {Ehle}, \& {Trinchieri}}]{pietsch04}{Pietsch}, W., {Misanovic}, Z., {Haberl}, F., {Hatzidimitriou}, D., {Ehle}, M.,  \& {Trinchieri}, G. 2004, \aap, 426, 11 [PMH04]
\bibitem[Pietsch et al.(2006)]{pietsch06} Pietsch, W., Haberl, F., Sasaki, M., Gaetz, T.~J., Plucinsky, P.~P., Ghavamian, P., Long, K.~S., \& Pannuti, T.~G.\ 2006, \apj, 646, 420 
\bibitem[Pietsch et al.(2009)]{pietsch09} Pietsch, W., et al.\ 2009, \apj, 694, 449 
\bibitem[Plucinsky et al.(2008)]{ppp08} Plucinsky, P.~P., et al.\ 2008, \apjs, 174, 366 
\bibitem[Prestwich et al.(2009)]{prest09} Prestwich, A.~H., Kilgard, R.~E., Primini, F., McDowell, J.~C., \& Zezas, A.\ 2009, \apj, 705, 1632 
\bibitem[Robitaille \& Whitney(2010)]{robitaille2010} Robitaille, T.~P., \& Whitney, B.~A.\ 2010, \apjl, 710, L11 
\bibitem[Rosolowsky \& Simon(2008)]{rosolowsky2010} Rosolowsky, E., \& Simon, J.~D.\ 2008, \apj, 675, 1213 
\bibitem[Smith et al.(1978)]{smith1978} Smith, L.~F., Biermann, P., \& Mezger, P.~G.\ 1978, \aap, 66, 65 \bibitem[Smith et al.(2001)]{smith01} Smith, R.~K., Brickhouse, N.~S., Liedahl, D.~A., \& Raymond, J.~C.\ 2001, \apjl, 556, L91 
\bibitem[Sarajedini \& Mancone(2007)]{sara07} Sarajedini, A., \& Mancone, C.~L.\ 2007, \aj, 134, 447 [SM07]
\bibitem[Schulman \& Bregman(1995)]{schube95} Schulman, E., \& Bregman, J.~N.\ 1995, \apj, 441, 568 
\bibitem[Supper et al.(2001)]{supper01} Supper, R., Hasinger, G., Lewin, W.~H.~G., Magnier, E.~A., van Paradijs, J., Pietsch, W., Read, A.~M., \& Tr{\"u}mper, J.\ 2001, \aap, 373, 63 
\bibitem[Trinchieri et al.(1988)]{trine88} Trinchieri, G., Fabbiano, G., \& Peres, G.\ 1988, \apj, 325, 531 
\bibitem[T{\"u}llmann et al.(2008)]{tull08} T{\"u}llmann, R., et al.\ 2008, \apj, 685, 919 
\bibitem[T{\"u}llmann et al.(2009)]{tull09} T{\"u}llmann, R., et al.\ 2009, \apj, 707, 1361
\bibitem[Verley et al.(2009)]{verley2009} Verley, S., Corbelli, E., Giovanardi, C., \& Hunt, L.~K.\ 2009, \aap, 493, 453
\bibitem[Weng et al.(2009)]{wewa09} Weng, S.-S., Wang, J.-X., Gu, W.-M., \& Lu, J.-F.\ 2009, \pasj, 61, 1287 
\bibitem[Williams et al.(2008)]{ben08} Williams, B.~F., et al.\ 2008, \apj, 680, 1120 [W08]
\bibitem[Wilms et al.(2000)]{wilms00} Wilms, J., Allen, A., \& McCray, R.\ 2000, \apj, 542, 914 
\bibitem[Zaritsky et al.(1989)]{zar89} Zaritsky, D., Elston, R., \& Hill, J.~M.\ 1989, \aj, 97, 97 
\bibitem[Zezas et al.(2007)]{zezas07} Zezas, A., Fabbiano, G., Baldi, A., Schweizer, F., King, A.~R., Rots, A.~H., \& Ponman, T.~J.\ 2007, \apj, 661, 135 
\bibitem[Zloczewski et al.(2008)]{zloc08} Zloczewski, K., Kaluzny, J., \& Hartman, J.\ 2008, Acta Astronomica, 58, 23 [ZKH08]
\end{thebibliography}
\end{document}